%% file: sauter-schwinger_in_semiconductors.tex
\documentclass%
[aps,prb,reprint,showpacs,longbibliography,nobibnotes,amsfonts,amssymb,amsmath]%
{revtex4-1}

\setcitestyle{numbers,square}

\usepackage[T1]{fontenc}
\usepackage{booktabs} 
\usepackage{mathtools} 
\usepackage{graphicx}
\usepackage{bm} 
\usepackage[justification=centerlast]{caption}
\usepackage[listofformat=parens]{subfig}
\usepackage{braket} 

\makeatletter

\usepackage[pdfauthor={Malte F. Linder, Axel Lorke, and Ralf Sch\"{u}tzhold},
unicode=true,pdfusetitle,bookmarks=true,bookmarksnumbered=false,
bookmarksopen=true,bookmarksopenlevel=1,breaklinks=false,pdfborder={0 0 1},
backref=false,colorlinks=false]{hyperref}

\makeatother

\newcommand{\I}{\mathrm{i}} 
\newcommand{\E}{\mathrm{e}} 

\newcommand{\D}{\mathrm{d}}               
\renewcommand{\Re}{\operatorname{Re}}     
\DeclareMathOperator{\DiracDelta}{\delta} 
\DeclareMathOperator{\order}{\mathcal{O}} 

\newcommand{\spi}[1]{\underline{#1}} 

\newcommand{\energy}{\mathcal{E}} 
\newcommand{\Ef}{E}           
\newcommand{\Phim}{\tilde{\Phi}} 
\newcommand{\InstAct}{\mathcal{A}} 

\newcommand{\cvac}{c}                 
\newcommand{\epsilonvac}{\varepsilon_0} 

\newcommand{\LC}{\ell} 
\newcommand{\Eg}{\energy_{g}} 
\newcommand{\EF}{\energy_{F}} 
\newcommand{\lambdaC}{\lambda_{C}} 
\newcommand{\kc}{K} 

\newcommand{\SC}[2]{#1^{\mathrm{#2}}} 

\newcommand{\ceff}{\cvac_\star} 
\newcommand{\meff}{m_\star} 
\newcommand{\meffc}{m_{\star,e}} 
\newcommand{\meffv}{m_{\star,h}} 
\newcommand{\qeff}{q_\star} 
\newcommand{\Aeff}{A_\star} 

\newcommand{\HOP}{\hat{H}} 
\newcommand{\HOPvar}{\hat{\mathcal{H}}} 
\newcommand{\HDirac}{\HOP_{D}} 
\newcommand{\HSC}{\HOP_{s}} 
\newcommand{\HSCfull}{\HSC^{\mathrm{full}}} 

\newcommand{\DFO}{\hat{\Psi}} 
\newcommand{\DFOm}{\hat{\widetilde{\Psi}}} 
\newcommand{\RDFO}{\hat{\Upsilon}} 

\newcommand{\SCFO}{\hat{\psi}} 
\newcommand{\AOP}{\hat{a}} 
\newcommand{\RAOP}{\hat{b}} 

\newcommand{\vbi}{-} 
\newcommand{\cbi}{+} 
\newcommand{\covbi}{\pm} 
\newcommand{\vocbi}{\mp} 

\newcommand{\braketcell}[1]{\braket{#1}_{u}}
\newcommand{\Braketcell}[1]{\Braket{#1}_{u}}

\newcommand{\Ecrit}[1][]{\Ef_{\mathrm{crit}}^{\mathrm{#1}}} 
\newcommand{\Estatic}{\Ef_{\mathrm{stat}}} 
\newcommand{\ESauter}{\Ef_{\mathrm{Sauter}}} 
\newcommand{\Ecos}{\Ef_{\mathrm{wave}}} 
\newcommand{\Ecoscrit}{\Ecos^{\mathrm{crit}}} 
\newcommand{\KP}{\gamma} 
\newcommand{\KPcrit}{\KP^{\mathrm{crit}}} 
\newcommand{\omegacrit}{\omega_{\mathrm{crit}}} 
\newcommand{\Icrit}{I_{\mathrm{crit}}} 
\newcommand{\Etwocrit}{\Ef_{2}^{\mathrm{crit}}} 

\newcommand{\ti}{\tau} 
\newcommand{\xT}{x^{\star}} 

\newcommand{\PMD}{M} 
\newcommand{\PMSC}{\mathcal{M}} 
\newcommand{\ES}{\Ecrit[QED]} 
\newcommand{\Fext}{F_{\mathrm{ext}}} 
\newcommand{\FCoul}{F_{\mathrm{Coulomb}}} 
\newcommand{\alphaQED}{\alpha_{\mathrm{QED}}} 

\begin{document}
\title{Analog Sauter--Schwinger effect in semiconductors for spacetime-dependent
fields}

\author{Malte F.\ \surname{Linder}}
\author{Axel \surname{Lorke}}
\author{Ralf \surname{Sch\"{u}tzhold}}
\email{ralf.schuetzhold@uni-due.de}

\affiliation{Fakult\"{a}t~f\"{u}r~Physik, Universit\"{a}t~Duisburg-Essen,
Lotharstr.~1, 47057~Duisburg, Germany}

\date{January 16, 2018}

\pacs{12.20.-m, 77.22.Jp, 11.15.-q}


\begin{abstract}
The Sauter--Schwinger effect predicts the creation of elec\-tron--pos\-i\-tron
pairs out of the quantum vacuum via tunneling induced by a strong electric
field.
Unfortunately, as the required field strength is extremely large, this
fundamental prediction of quantum field theory has not been verified
experimentally yet.
Here, we study under which conditions and approximations the interband
tunneling in suitable semiconductors could be effectively governed by the same
(Dirac) Hamiltonian, especially for electric fields which depend
on space and time.
This quantitative analogy would allow us to test some of the predictions
(such as the dynamically assisted Sauter--Schwinger effect) in this area by
means of these laboratory analogs.
\end{abstract}

\maketitle

\section{Introduction}

The are several fundamental predictions of quantum field theory which have so 
far resisted a direct experimental verification.
One of the most prominent examples is the Sauter--Schwinger effect~%
\cite{Sauter1931,Sauter1932,Heisenberg+Euler1936,Schwinger1951}
predicting the creation of elec\-tron--pos\-i\-tron pairs out of the quantum
vacuum via tunneling.
For a constant electric field $\Ef$, the associated pair-cre\-a\-tion
probability behaves as
\begin{equation}
P_{e^{+}e^{-}}\propto\E^{-\pi\ES/\Ef}
\label{eq:Schwinger-rate}
\end{equation}
and is thus exponentially suppressed for fields $\Ef$ well below the critical
field
\begin{equation}
\label{eq:ES}
\ES=\frac{m^2 \cvac^3}{\hbar q} \approx 1.3\times 10^{18}\,\mathrm{\frac{V}{m}}
\end{equation}
(often denoted by $\Ef_S$ in the literature), where $m$ is the electron mass
and $q>0$ the elementary charge.
The fact that expression~\eqref{eq:Schwinger-rate} does not admit a Taylor
expansion in $q$ already indicates that this is a nonperturbative
effect, which renders calculations intrinsically difficult.
Nevertheless, apart from the con\-stant-field case above, it is possible to
derive the pair-cre\-a\-tion probability for several scenarios with varying
fields.
For example, a temporal Sauter pulse of the form
$\Ef(t)=\Ef_0/\cosh^2(\omega t)$ does also facilitate an exact solution of the
Dirac equation (see, e.g., Ref.~\cite{Popov1972}).
In this situation, the absolute value of the exponent~\eqref{eq:Schwinger-rate}
is reduced and thus the probability enhanced.
Conversely, for a spatial Sauter profile $\Ef(x)=\Ef_0/\cosh^2(kx)$, the
absolute value of the exponent increases, leading to a suppression of the
pair-cre\-a\-tion probability.

As another interesting case, the superposition of a constant (or slowly varying)
strong field with a weaker time-de\-pen\-dent field can result in an enhancement
of the probability: the dynamically assisted Sauter--Schwinger
effect~\cite{Schuetzhold+Gies+Dunne2008}.
The dependence of this effect on the shape of the weaker time-de\-pen\-dent
field and the momentum of the created electrons and positrons has been studied
in Refs.~\cite{Orthaber+Hebenstreit+Alkofer2011,Fey+Schuetzhold2012,%
Kohlfuerst+al2013,Linder+al2015}, for example.
If the strong field is not constant but spatially varying (such as a spatial
Sauter profile), there is an interesting interplay or competition between the
spatial dependence of the stronger field and the temporal dependence of the
weaker field; see Ref.~\cite{Schneider+Schuetzhold2016-1}.

Most unfortunately, because the critical field strength~\eqref{eq:ES} is so
large, these nonperturbative phenomena have not been observed yet, and thus it
was not possible to test the various predictions mentioned above experimentally.
This motivates the quest for other, experimentally more
accessible, laboratory systems which display analogous effects,
ideally governed by the same Hamiltonian (under appropriate approximations)
and thus the same equations of motion.
To use the famous quote by R.~Feynman:
``\emph{The same equations have the same solutions.}''
Due to their high degree of experimental control, one such option are ultracold
atoms in optical lattices; see also Refs.~\cite{Szpak+Schuetzhold2011,%
Szpak+Schuetzhold2012,Kasper+al2016}.
Other possible options include graphene~\cite{Allor+Cohen+McGady2008,%
Dora+Moessner2010,Akal+al2016} and trapped ions~\cite{Martinez+al2016}.
In the following, we study interband tunneling in semiconductors as another
promising example.
Note that the qualitative analogy between Landau--Zener tunneling in
semiconductors~\cite{Landau1932,Zener1932,Zener1934} and the Sauter--Schwinger
effect in the case of a constant electric field has already been discussed in,
e.g., Refs.~\cite{Hrivnak1993,Rau+Mueller1996,Zawadzki2005-1,Oka+Aoki2005,%
Smolyansky+Tarakanov+Bonitz2009,Dora+Moessner2010}.
Here, the goal is to derive a quantitative analogy (in the spirit of Feynman)
and to specify the underlying approximations and assumptions, with special
emphasis on fields depending on time (see also
Ref.~\cite{Smolyansky+Tarakanov+Bonitz2009}) and space (as motivated above).
The use of these analogies is twofold:
On the one hand, they allow us to test the above predictions by means of
laboratory analogs, which are easier to access experimentally, and, on the
other hand, they help us to understand the physics of these laboratory systems
better.

\boldmath
\section{\label{sec:A(t)-case}Time-dependent case $\Ef=\Ef(t)$}
\unboldmath
Let us start with the simpler case of a homogeneous and purely
time-de\-pen\-dent external electric field in 1+1 spacetime dimensions.
We choose to describe the external electric field in temporal gauge
$\Ef(t)=\dot{A}(t)$ with the one-com\-po\-nent vector potential $A(t)$.
This potential couples to the electron momentum operator via the covariant
derivative $\partial_{x}+\I qA(t)$ ($\cvac=\epsilonvac=\hbar=1$ in the
following, unless otherwise stated).

The many-body Dirac Hamiltonian can be written as
\begin{equation}
\HDirac(t)=\intop_{-\infty}^{\infty}\spi{\DFO}^{\dagger}
\left\{ \left[-\I\partial_{x}+qA(t)\right]\sigma_{x}
+m\sigma_{z}\right\} \spi{\DFO}\,\D x\text{.}
\label{eq:A(t)-HDirac-pos-space}
\end{equation}
This form is obtained by expressing the Dirac matrices in terms of Pauli
matrices via $\gamma^{0}=\sigma_{z}$ and $\gamma^{1}=\I\sigma_{y}$.
The field operator consequently has two components,
$\spi{\DFO}(t,x)=\bigl(\DFO_{+}(t,x),\DFO_{-}(t,x)\bigr)$, which corresponds
to the absence of spin in 1+1 dimensions.

We transform this Hamiltonian to momentum space by inserting the spatial
Fourier transform
\begin{equation}
\spi{\DFO}(t,x)=\frac{1}{\sqrt{2\pi}}\intop_{-\infty}^{\infty}
\spi{\DFOm}(t,k)\E^{\I kx}\,\D k
\label{eq:DFO-Fourier-trafo}
\end{equation}
of the field operator.
The result reads
\begin{equation}
\HDirac(t)=\!\intop_{-\infty}^{\infty}\spi{\DFOm}^{\dagger}(t,k)
\begin{pmatrix}m & k+qA(t)\\
k+qA(t) & -m\end{pmatrix}\spi{\DFOm}(t,k)\,\D k\text{.}
\label{eq:A(t)-HDirac-mom-space}
\end{equation}

The next step is to derive the crys\-tal-mo\-men\-tum representation
of the Hamiltonian for electrons in a semiconductor which is exposed
to the same external electric field.
This semiconductor Hamiltonian can then be compared to the Dirac
Hamiltonian~\eqref{eq:A(t)-HDirac-mom-space}.

\subsection{\label{subsec:A(t)-SC-model}Two-band semiconductor model}

A direct, quantitative analogy between Dirac's theory and electrons
in a semiconductor can only exist if the considered semiconductor
electrons can only occupy two adjacent energy bands: the
higher (lower) band then corresponds to the positive (negative) relativistic
continuum.
In the ground state (no external field and zero temperature),
the lower band must be completely filled with electrons (analog of
the Dirac sea), while the upper band must be empty.
This is precisely the case if we restrict the semiconductor model
to the valence band and the conduction band only.
Our starting point is the well-known
Kane model~\cite{Kane1957}, but we only include the light-hole
valence band in our theory and neglect the heavy-hole band (since
lighter particles are more likely to be excited via the pair-cre\-a\-tion
mechanism we are interested in) and the split-off valence band, which
is energetically lowered due to spin--orbit interaction; see, e.g.,
Refs.~\cite{Zawadzki+Lax1966,Aronov+Pikus1967-1,Zawadzki+Rusin2011,%
Zawadzki2013}, which also employ and describe this model.

Let us start with the basic Hamiltonian.
Since the possible electron
group velocities within the valence and conduction bands of typical
semiconductors are far below the vacuum speed of light, we may describe
the semiconductor electrons with the nonrelativistic Schr\"{o}dinger
equation.
The Bloch electrons, which we
are interested in, are subject to the lat\-tice-peri\-od\-ic potential
$V(x)$ of the ion cores.
We denote the lattice constant by $\LC$, so the potential satisfies
$V(x+\LC)=V(x)$.
For simplicity, we neglect elec\-tron--elec\-tron interactions; see
Sec.~\ref{sec:outlook} below.
The Hamiltonian of the Bloch electrons in the external field
$\Ef(t)=\dot{A}(t)$ thus reads as
\begin{equation}
\HSCfull(t)=\intop_{-\infty}^{\infty}\SCFO^{\dagger}\left\{
\frac{[-\I\partial_{x}+qA(t)]^{2}}{2m}+V(x)\right\} \SCFO\,\D x\text{,}
\label{eq:A(t)-HSCfull}
\end{equation}
where $\SCFO(t,x)$ is the scalar electron field operator.

Note that the quadratic $A$ term in this Hamiltonian can be absorbed via a
suitable gauge transformation (see Appendix~\ref{app:A-squared-absorption}),
so we may consider the simplified Hamiltonian
\begin{equation}
\HSCfull(t)=\intop_{-\infty}^{\infty}\SCFO^{\dagger}\left[
-\frac{\partial_{x}^{2}}{2m}+V(x)+\frac{qA(t)}{m}(-\I\partial_{x})\right]
\SCFO\,\D x
\label{eq:A(t)-HSCfull-simplified}
\end{equation}
instead.

For the derivation of the two-band model, we restrict the Hamiltonian
$\HSCfull$ to valence- and con\-duc\-tion-band electrons only.
This should be a good approximation for analogs of the Sauter--Schwinger
effect since an excitation of a va\-lence-band electron into the
conduction band is associated with a lower energy difference than
any other possible transition in the (initial) ground state.
Larger energy differences lead to exponential suppression in the context
of nonperturbative pair creation, so the two-band model should reproduce
the lead\-ing-order pair-cre\-a\-tion probability in the initial ground
state correctly.

We apply the two-band approximation by assuming that only the valence
and conduction Bloch bands contribute to the field operator:
\begin{equation}
\SCFO(t,x)\approx\intop_{\mathclap{-\pi/\LC}}^{\mathclap{\pi/\LC}}\AOP_{\vbi}
(t,\kc)f_{\vbi}(\kc,x)+\AOP_{\cbi}(t,\kc)f_{\cbi}(\kc,x)\,\D\kc\text{.}
\label{eq:Two-band-field-op}
\end{equation}
In this equation, the functions $f_{n}(\kc,x)=\braket{x|n,\kc}$ are
the po\-si\-tion-space representations of the Bloch states $\ket{n,\kc}$
in the unperturbed semiconductor crystal {[}$A(t)=0${]}.
The band index $-$ ($+$) denotes the valence (conduction) band.
There is one independent Bloch state per band for each quasimomentum $\kc$ in
the first Brillouin zone, which is the range $(-\pi/\LC,\pi/\LC]$.
Hence, our field operator~\eqref{eq:Two-band-field-op} is per assumption
a linear combination of all Bloch states in the valence and the conduction
bands at each instant of time.
The time-de\-pen\-dent ``coefficients,'' which are in fact operators,
$\AOP_{\covbi}(t,\kc)$, are instantaneous annihilation operators for electrons
in the corresponding Bloch states $\ket{\covbi,\kc}$. For this statement to
hold, the Bloch states must be normalized, so that they obey the orthonormality
relation
\begin{multline}
\braket{n,\kc|n',\kc'}=\intop_{-\infty}^{\infty}f_{n}^{\ast}(\kc,x)
f_{n'}(\kc',x)\,\D x\\ {}=\delta_{nn'}\DiracDelta(\kc'-\kc)\text{.}
\label{eq:BW-orthonormality-rel}
\end{multline}
We use the convention
\begin{equation}
f_{n}(\kc,x)=\E^{\I\kc x}u_{n}(\kc,x)
\label{eq:BW-form}
\end{equation}
throughout this paper, so our lat\-tice-peri\-od\-ic Bloch factors
$u_{n}(\kc,x)$ are orthonormalized (at a fixed $\kc$) according
to the unit-cell Bloch-fac\-tor scalar product
\begin{equation}
\braketcell{n,\kc|n',\kc}=\frac{2\pi}{\LC}\intop_{0}^{\LC}u_{n}^{\ast}(\kc,x)
u_{n'}(\kc,x)\,\D x=\delta_{nn'}\text{.}
\label{eq:Bloch-factor-orthonormalization}
\end{equation}

Inserting the approximation~\eqref{eq:Two-band-field-op} into the full
Hamiltonian~\eqref{eq:A(t)-HSCfull-simplified} yields the two-band
semiconductor Hamiltonian $\HSC$, which neglects the dynamics of
all other Bloch bands.
In the calculation of $\HSC$, we use the fact that Bloch waves satisfy
the energy eigenvalue equation
\begin{equation}
\left[-\frac{\partial_{x}^{2}}{2m}+V(x)\right]f_{n}(\kc,x)=
\energy_{n}(\kc)f_{n}(\kc,x)\text{.}
\label{eq:BW-eigenvalue-eq}
\end{equation}
Furthermore, the Bloch-wave momentum matrix elements
$\braket{n,\kc|-\I\partial_{x}|n',\kc'}$ (also known as optical matrix elements)
appear in the new Hamiltonian.
It is well known that these matrix elements vanish unless $\kc=\kc'$ (see, e.g.,
Ref.~\cite{Gu+Kwong+Binder2013}; a proof of this important theorem is given in
Appendix~\ref{app:mom-matrix}).
There are thus three independent
momentum matrix elements in the two-band model for each $\kc$: the interband
element $\kappa$ is given implicitly by
$\braket{\vbi,\kc|-\I\partial_{x}|\cbi,\kc'}=\kappa(\kc)\DiracDelta(\kc'-\kc)$
{[}cf.\ Eq.~\eqref{eq:BW-mom-matrix-elem}{]} and can be written
\begin{equation}
\kappa(\kc)=\braketcell{\vbi,\kc|-\I\partial_{x}|\cbi,\kc}
\label{eq:kappa-def}
\end{equation}
with the product defined in Eq.~\eqref{eq:Bloch-factor-orthonormalization}.
This quantity is complex in general; however, we define the global
phases of the Bloch bands in a way such that the value
$\kappa_{0}=\kappa(0)$ is real and positive: $\kappa_{0}>0$.
The two intraband elements are related to the group velocities
$v_{\covbi}(\kc)=\D\energy_{\covbi}(\kc)/\D\kc$ via
\begin{equation}
\braket{\covbi,\kc|-\I\partial_{x}|\covbi,\kc'}=mv_{\covbi}(\kc)
\DiracDelta(\kc'-\kc)\text{;}
\label{eq:intraband-mom-matrix-elem}
\end{equation}
see, e.g., Refs.~\cite{Ashcroft+Mermin2008,Zawadzki2013}.

The resulting two-band Hamiltonian in crys\-tal-mo\-men\-tum space reads
\begin{multline}
\HSC(t)\\ {}=\intop_{\mathclap{-\pi/\LC}}^{\mathclap{\pi/\LC}}
\spi{\AOP}^{\dagger}(t,\kc)
\begin{pmatrix}\energy_{\cbi}+qAv_{\cbi} & \frac{qA}{m}\kappa^{\ast}\\
\frac{qA}{m}\kappa & \energy_{\vbi}+qAv_{\vbi}\end{pmatrix}
\spi{\AOP}(t,\kc)\,\D\kc
\label{eq:A(t)-HSC}
\end{multline}
(we have omitted to write explicitly the dependencies of the quantities in the
matrix here) with
\begin{equation}
\spi{\AOP}(t,\kc)=\begin{pmatrix}\AOP_{\cbi}(t,\kc)\\
\AOP_{\vbi}(t,\kc)\end{pmatrix}\text{.}
\end{equation}

Note that this Hamiltonian as well as the Dirac
Hamiltonian~\eqref{eq:A(t)-HDirac-mom-space} have the form
$\HOP(t)=\int\HOPvar(t,k)\,\D k$, which means that each $k$ mode evolves
independently, and $k$ (or $\kc$ in the semiconductor case) is thus a conserved
quantity as expected in a purely time-de\-pen\-dent potential.

\subsection{\label{subsec:A(t)-H-diagonalization}%
Diagonalization of the Hamiltonians}

In order to bring both Hamiltonians, $\HDirac$ and $\HSC$, into
the same form, so that we can compare them, we diagonalize the $2\times2$
matrices in the Hamiltonians.
To this end, we transform (``rotate'') the mo\-men\-tum-space field operators
$\spi{\DFOm}(t,k)$ (Dirac case) and the Bloch-elec\-tron operators
$\spi{\AOP}(t,\kc)$ (semiconductor) to operators corresponding to the
instantaneous energy eigenstates, respectively.

In the Dirac case, the transformed field operators read as
\begin{equation}
\spi{\RDFO}(t,k)=\frac{1}{\sqrt{1+d^{2}(t,k)}}\begin{pmatrix}1 & d(t,k)\\
-d(t,k) & 1\end{pmatrix}\spi{\DFOm}(t,k)
\label{eq:A(t)-DFO-rotated}
\end{equation}
with the abbreviations
\begin{equation}
d(t,k) = \frac{k+qA(t)}{m+\Omega(t,k)}
\label{eq:A(t)-Dirac-aux-d}
\end{equation}
and
\begin{equation}
\Omega(t,k)=\sqrt{m^{2}+[k+qA(t)]^{2}}\text{.}
\label{eq:A(t)-Dirac-Omega}
\end{equation}
Note that Eq.~\eqref{eq:A(t)-DFO-rotated} describes a unitary relation,
which is also a Bogoliubov transformation, so the two components of
$\spi{\RDFO}$ obey the canonical anticommutation relations.
In terms of these field operators, the Dirac
Hamiltonian~\eqref{eq:A(t)-HDirac-mom-space} assumes the diagonal form
\begin{equation}
\HDirac(t)=\intop_{-\infty}^{\infty}\spi{\RDFO}^{\dagger}(t,k)
\begin{pmatrix}\Omega(t,k) & 0\\ 0 & -\Omega(t,k)\end{pmatrix}
\spi{\RDFO}(t,k)\,\D k\text{.}
\label{eq:A(t)-HDirac-diag}
\end{equation}

Before we diagonalize the matrix in the semiconductor
Hamiltonian~\eqref{eq:A(t)-HSC}, we want to make its diagonal elements
symmetric like in the Dirac case, in which the original diagonal elements
in Eq.~\eqref{eq:A(t)-HDirac-mom-space} are $\pm m$.
In order to do this, we rewrite the Hamiltonian as
\begingroup
\renewcommand*{\arraystretch}{1.5}
\begin{align}
 & \HSC(t) \nonumber \\
={} & \intop_{\mathclap{-\pi/\LC}}^{\mathclap{\pi/\LC}}
\spi{\AOP}^{\dagger}(t,\kc)
\begin{pmatrix}
\frac{\Delta\energy+qA\Delta v}{2} & \frac{qA}{m}\kappa^{\ast}\\
\frac{qA}{m}\kappa                 & -\frac{\Delta\energy+qA\Delta v}{2}
\end{pmatrix}
\spi{\AOP}(t,\kc)\,\D\kc \nonumber \\
 & {}+\intop_{\mathclap{-\pi/\LC}}^{\mathclap{\pi/\LC}}
\frac{\energy_{\cbi}(\kc)+\energy_{\vbi}(\kc)
+qA(t)[v_{\cbi}(\kc)+v_{\vbi}(\kc)]}{2} \nonumber \\
 & \quad\times\Big[\underbrace{\AOP_{\cbi}^{\dagger}(t,\kc)\AOP_{\cbi}(t,\kc)
 +\AOP_{\vbi}^{\dagger}(t,\kc)\AOP_{\vbi}(t,\kc)}
 _{=1\,\text{for all}\,t\,\text{and}\,\kc} \Big]\,\D\kc \text{.}
\label{eq:A(t)-HSC-intermed}
\end{align}
\endgroup
In this equation, we have introduced the ($\kc$-de\-pen\-dent) band-en\-er\-gy
difference $\Delta\energy(\kc)=\energy_{\cbi}(\kc)-\energy_{\vbi}(\kc)$ and
the group-ve\-loc\-i\-ty difference $\Delta v(\kc)=v_{\cbi}(\kc)-v_{\vbi}(\kc)$.
Since $\kc$ is a conserved quantity,
$\spi{\AOP}^{\dagger}(t,\kc)\spi{\AOP}(t,\kc)$ must always be $1$ because there
is exactly one electron per $\kc$ value in our two-band model, and the electron
for a given $\kc$ must be either in the conduction band or in the valence band
at each point in time.
The second $\kc$ integral in the Hamiltonian~\eqref{eq:A(t)-HSC-intermed}
therefore yields a time-de\-pen\-dent constant, which can be eliminated by a
gauge transformation on the scalar potential again, as described in
Appendix~\ref{app:A-squared-absorption}.

The Bogoliubov transformation which diagonalizes the redefined semiconductor
Hamiltonian {[}first $\kc$ integral in Eq.~\eqref{eq:A(t)-HSC-intermed}{]} has
the same form as in the Dirac case {[}complex version of
Eq.~\eqref{eq:A(t)-DFO-rotated}{]},
\begin{equation}
\spi{\RAOP}(t,\kc)=\frac{1}{\sqrt{1+|\mathfrak{d}(t,\kc)|^{2}}}
\begin{pmatrix}1 & \mathfrak{d}^{\ast}(t,\kc)\\ -\mathfrak{d}(t,\kc) & 1
\end{pmatrix}\spi{\AOP}(t,\kc)\text{,}
\end{equation}
but with different auxiliary functions
\begin{equation}
\mathfrak{d}(t,\kc)=\frac{qA(t)\kappa(\kc)/m}
{[\Delta\energy(\kc)+qA(t)\Delta v(\kc)]/2+\varOmega(t,\kc)}
\end{equation}
and
\begin{multline}
\varOmega(t,\kc)\\ {}=\sqrt{\left[\frac{\Delta\energy(\kc)+qA(t)\Delta v(\kc)}
{2}\right]^{2}+\left[\frac{qA(t)|\kappa(\kc)|}{m}\right]^{2}}\text{,}
\label{eq:A(t)-SC-Omega}
\end{multline}
so the resulting Hamiltonian reads as
\begin{equation}
\HSC(t)=\intop_{\mathclap{-\pi/\LC}}^{\mathclap{\pi/\LC}}
\spi{\RAOP}^{\dagger}(t,\kc)
\begin{pmatrix}\varOmega(t,\kc) & 0\\ 0 & -\varOmega(t,\kc)
\end{pmatrix}\spi{\RAOP}(t,\kc)\,\D\kc\text{.}
\label{eq:A(t)-HSC-diag}
\end{equation}

Now that we have derived the diagonal forms of both Hamiltonians, their physical
differences including scales and dependence on conserved
\mbox{(quasi)}mo\-men\-tum are encoded in the instantaneous energy eigenvalues
$\Omega$ and $\varOmega$.

\boldmath
\subsection{\label{subsec:analogy-at-band-gap}%
Analogy between the modes $k=0$ and $\kc=0$}
\unboldmath

Let us start to point out the quantitative analogy between the two
Hamiltonians, $\HDirac$~\eqref{eq:A(t)-HDirac-diag} and
$\HSC$~\eqref{eq:A(t)-HSC-diag}, at the \mbox{(quasi)}mo\-men\-tum-space points
$k=\kc=0$.
We assume for the moment that there is no electric field ($A=0$).

\begin{figure}
\subfloat[%
\label{fig:disp-rel-Dirac}%
Relativistic dispersion relation.]{\includegraphics{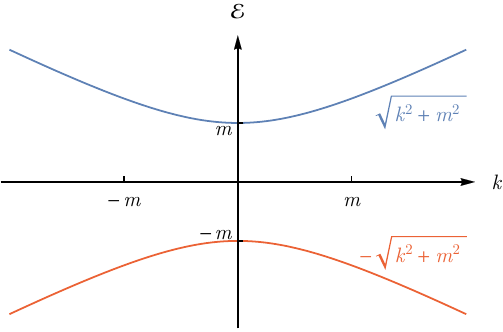}}

\subfloat[%
\label{fig:disp-rel-SC}%
Reduced zone scheme of a two-band semiconductor.]{\includegraphics{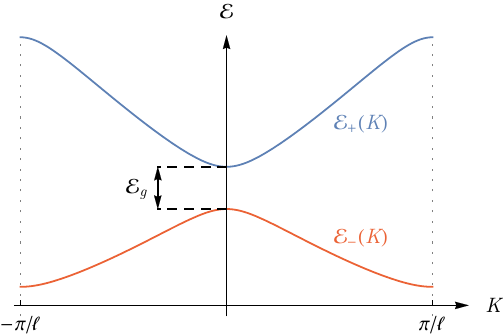}}

\caption{\label{fig:disp-rels}%
Electron dispersion relations in the two systems
under consideration, without an external electric field ($A=0$).
(\protect\subref*{fig:disp-rel-Dirac})~%
Dirac case: the two branches of the relativistic
en\-er\-gy--mo\-men\-tum relation.
(\protect\subref*{fig:disp-rel-SC})~%
Semiconductor case: example for an electronic two-band structure in the first
Brillouin zone (reduced zone scheme).
We assume throughout this paper that the semiconductor has a direct band gap at
the center of the Brillouin zone, measuring $\Eg=\Delta\energy(0)$.}
\end{figure}

The energy bands in the Dirac case are the two square roots of the relativistic
en\-ergy--momen\-tum relation; see Fig.~\subref{fig:disp-rel-Dirac}.
The mode $k=0$ thus coincides with the minimal mass gap $2m$ in the absence of
an external field.

The exact shapes of the valence band and the conduction band in the
semiconductor, $\energy_{\covbi}(\kc)$, are not fixed but depend on the
periodic potential $V(x)$; however, we make the following assumptions about
the semiconductor band structure, which shall be satisfied in the
remainder of this paper:
\begin{itemize}
\item no band crossing {[}$\Delta\energy(\kc)>0$ for each $\kc${]} and
\item a direct band gap at the center $\kc=0$ of the Brillouin zone; that is,
$\Eg=\Delta\energy(0)$ is the minimal value of $\Delta\energy(\kc)$.
\end{itemize}
An example for such a band structure is plotted in
Fig.~\subref{fig:disp-rel-SC}.


Now, we reintroduce the electric field and compare the instantaneous energy
eigenvalues in Eqs.~\eqref{eq:A(t)-Dirac-Omega} and~\eqref{eq:A(t)-SC-Omega}
at $k=\kc=0$ with each other.
In the Dirac case, we get
\begin{equation}
\Omega(t,0)=\sqrt{(m\cvac^{2})^{2}+[\cvac qA(t)]^{2}}
\label{eq:A(t)-k=00003D0-Dirac-Omega}
\end{equation}
with the speed of light written explicitly in this equation.
In the semiconductor case, we first note that both group velocities
{[}$\energy_{\covbi}(\kc)$ derivatives{]} vanish at $\kc=0$ and
thus also $\Delta v(0)=0$.
Comparing the resulting
$\varOmega(t,0)=\sqrt{(\Eg/2)^{2}+[qA(t)\kappa_{0}/m]^{2}}$ with
Eq.~\eqref{eq:A(t)-k=00003D0-Dirac-Omega}, we immediately see that the quantity
$\ceff=\kappa_{0}/m$ plays the role of an effective speed of light in the
semiconductor.

We also want to define a suitable effective mass $\meff$ such that
$\meff\ceff^{2}$ (the analog of the rest energy $m\cvac^{2}$ in
Dirac theory) produces the term $\Eg/2$ in $\varOmega(t,0)$ above.
Hence, we set
\begin{equation}
\meff=\frac{m^{2}\Eg}{2\kappa_{0}^{2}}\text{,}
\label{eq:meff-of-kappa}
\end{equation}
so we may write
\begin{equation}
\ceff=\sqrt{\frac{\Eg}{2\meff}}
\label{eq:ceff}
\end{equation}
and
\begin{equation}
\varOmega(t,0)=\sqrt{(\meff\ceff^{2})^{2}+[\ceff qA(t)]^{2}}\text{.}
\label{eq:A(t)-k=00003D0-SC-Omega}
\end{equation}
Comparing Eqs.~\eqref{eq:A(t)-k=00003D0-Dirac-Omega}
and~\eqref{eq:A(t)-k=00003D0-SC-Omega} shows that the Hamiltonians of both
systems are equivalent in the large-wave\-length limit $k=\kc=0$.
The semiconductor just exhibits different scales, which are given by the
material constants $\Eg$ and $\kappa_{0}$.
The same effective constants have also been found in
Refs.~\cite{Zawadzki+Lax1966,Aronov+Pikus1967-1}.

Note that we refer to the quantity~\eqref{eq:meff-of-kappa} as ``effective
mass'' because it allows us to write $\varOmega(t,0)$ in a way formally
equivalent to $\Omega(t,0)$ above.
Nevertheless, as we will see in the next subsection, $\meff$ is
indeed related to the parabolic en\-er\-gy-band curvatures in the semiconductor,
which is the usual notion of effective masses in this area of physics.

Another point to notice here is that we could also define an effective
elementary charge via $\cvac\qeff=q\kappa_0/m$ instead of the effective speed of
light~\eqref{eq:ceff} to make the analogy between the modes $k=\kc=0$ work (in
which case the effective electron mass in the semiconductor must be defined by
the equation $\meff\cvac^2=\Eg/2$).
Or, we can shift the factor in $\qeff$ into an effective vector potential
defined by $\qeff A(t)=q\Aeff(t)$.
The concept of an effective external potential in analogs of the
Sauter--Schwinger effect is known from ultracold atoms in optical lattices; see
Refs.~\cite{Szpak+Schuetzhold2011,Szpak+Schuetzhold2012}.
However, we will see in the next subsection that defining an effective speed of
light as above (and thus leaving the external potential and elementary charge
unchanged) is required to extend the analogy in the semiconductor to more modes
than just $k=\kc=0$.

\boldmath
\subsection{\label{subsec:analogy-gap-vicinity}%
Analogy for long-wave\-length modes}
\unboldmath

We now want to extend the analogy to all small
\mbox{(quasi)}mo\-men\-ta, which means $|k|\ll m$ in the Dirac case and
$|\kc|\ll\pi/\LC$ in the semiconductor case.
In this range, the dispersion curves in Fig.~\ref{fig:disp-rels} are
approximately parabolic in both cases.
This is also the range with the smallest energy difference between the bands,
and we consequently expect the corresponding modes to generate the dominant
contributions to the total pair-cre\-a\-tion yield via the Sauter--Schwinger
effect.
Note that if the vector potential vanishes in the \emph{in} and \emph{out}
states {[}i.e., $A(t \to \pm\infty)=0${]}, then the conserved quantities $k$ and
$\kc$ correspond to the initial \emph{and} the final kinetic
\mbox{(quasi)}mo\-men\-tum of the considered mode, respectively, so the
long-wave\-length modes can be identified with electron states close to the
minimal band gaps in Fig.~\ref{fig:disp-rels} in this case.
Given a particular external field $\Ef(t)$, we can always satisfy this condition
by letting $A(t)$ start at zero for $t \to -\infty$, and, if $A(t) \neq 0$ after
the field has been switched off (or has become very tiny), letting $A(t)$
approach zero again very slowly (adiabatically, such that this process does not
cause band transitions).

Concerning the analogy, let us start with the Dirac case again.
The Taylor expansion of $\Omega$ around $k=0$ reads
\begin{equation}
\Omega(t,k)=\Omega(t,0)+\frac{\cvac^{2}qA(t)}{\Omega(t,0)}k+\order(k^{2})
\label{eq:A(t)-gap-vic-Dirac-Omega}
\end{equation}
with $\cvac$ written explicitly.
In the sec\-ond-order term, the small quantity $(k/m)^2$ is suppressed
by the prefactor $1/\{2\,[1+q^{2}A^{2}(t)/m^{2}]^{3/2}\}\leq1/2$.
Hence, we will only consider the first orders of $k$ or $\kc$ in
$\Omega$ and $\varOmega$ when comparing the Hamiltonians for small
\mbox{(quasi)}mo\-men\-ta, and we ignore all high\-er-order terms.

In the semiconductor case, we get
\begin{align}
\varOmega(t,\kc)={} & \varOmega(t,0)+\frac{\kc}{\varOmega(t,0)}\nonumber \\
 & {}\times\left\{ \frac{\Delta\energy^{(1)}(0)+qA(t)\Delta v^{(1)}(0)}{4}\Eg
 \vphantom{+{\left[\frac{qA(t)}{m}\right]^{2}}\kappa_{0}\Re\kappa^{(1)}(0)}
 \right.\nonumber \\
 & \left.\vphantom{\frac{\Delta\energy^{(1)}(0)+qA(t)\Delta v^{(1)}(0)}{4}\Eg}
 \hspace*{1.5em}+{\left[\frac{qA(t)}{m}\right]^{2}}\kappa_{0}\Re\kappa^{(1)}(0)
 \right\} +\order(\kc^{2})\text{.}
\label{eq:A(t)-gap-vic-SC-Omega-v1}
\end{align}
Superscripts of the form ``$(n)$'' denote the $n$-th derivative with respect
to $\kc$.

In order to evaluate these derivatives at $\kc=0$, we first note
that the parabolic parts of the energy bands around the minimal gap
are usually written as (we arbitrarily locate the band gap symmetrically
around the zero energy level)
\begin{align}
\energy_{\cbi}(\kc) & =\frac{\Eg}{2}+\frac{\kc^{2}}{2\meffc}+
\order(\kc^{3})\text{,}\nonumber \\
\energy_{\vbi}(\kc) & =-\frac{\Eg}{2}-\frac{\kc^{2}}{2\meffv}+
\order(\kc^{3})\text{,}
\label{eq:A(t)-SC-parabolic-bands}
\end{align}
where $\meffc$ and $\meffv$ denote the (positive) effective masses
of con\-duc\-tion-band electrons and va\-lence-band holes in the
crystal.
These quantities can be calculated analytically from the
band structure by expanding the band energies in powers of $\kc$
up to the second order using $k\cdot p$ perturbation theory (see,
e.g., Ref.~\cite{Yu+Cardona2010}).
While doing so, we apply the two-band approximation again, which means that we
neglect contributions to $\meffc$ and $\meffv$ from other bands than the
valence band and the conduction band.
Within this model, we get the well-known relations
(cf.\ Ref.~\cite{Yu+Cardona2010})
\begin{align}
\frac{1}{\meffc} & =\frac{1}{m}+\frac{2\kappa_{0}^{2}}{m^{2}\Eg}
\text{,}\nonumber \\
\frac{1}{\meffv} & =-\frac{1}{m}+\frac{2\kappa_{0}^{2}}{m^{2}\Eg}
\end{align}
according to $k\cdot p$ perturbation theory.

By adding these two equations, we find that the effective mass
$\meff$~\eqref{eq:meff-of-kappa} defined in the previous subsection is given
by the harmonic mean of the effective charge-car\-ri\-er masses:
\begin{equation}
\meff=\frac{2}{\meffc^{-1}+\meffv^{-1}}\text{,}
\label{eq:meff-of-meffs}
\end{equation}
which equals twice the reduced mass.
This relation~\eqref{eq:meff-of-meffs} between our effective mass $\meff$,
which is related to $\kappa_{0}$ (off-diag\-o\-nal momentum matrix element)
via Eq.~\eqref{eq:meff-of-kappa}, and the parabolic curvatures of the energy
bands is essential for extending the analogy at $\kc=0$ to a neighborhood of
this point with the same effective physical constants, $\meff$ and $\ceff$,
as before.
Note that we are not required to assume $\meffc=\meffv$
here in the time-de\-pen\-dent case in order to draw the analogy.

If we had defined an effective elementary charge $\qeff$ instead of $\ceff$ (as
mentioned in the previous subsection), Eq.~\eqref{eq:meff-of-meffs} would not be
valid since we would have a different $\meff$ {[}given by $\Eg/(2\cvac^2)${]}
then.
For this reason, the analogy would not work for nonzero
\mbox{(quasi)}mo\-men\-ta.

Returning to the $\kc$ derivatives in Eq.~\eqref{eq:A(t)-gap-vic-SC-Omega-v1},
we may utilize Eq.~\eqref{eq:meff-of-meffs} to write the en\-er\-gy-band
difference in the semiconductor as
$\Delta\energy(\kc)=\Eg+\kc^{2}/\meff+\order(\kc^{3})$.
As expected at an extremum, we get $\Delta\energy^{(1)}(0)=0$.
The group velocity difference $\Delta v$ is given by the first $\kc$
derivative of the energy difference, so we get
$\Delta v^{(1)}(0)=\Delta\energy^{(2)}(0)=2/\meff$.
The quadratic $A$ term vanishes since $\kappa^{(1)}(0)=0$ (see
Appendix~\ref{app:kappa-Taylor-exp} for the calculation).
All in all, we arrive at
\begin{align}
\varOmega(t,\kc) & =\varOmega(t,0)+\frac{\Eg qA(t)/(2\meff)}{\varOmega(t,0)}
\kc+\order(\kc^{2})\nonumber \\
 & \overset{\mathllap{\text{Eq.\,\eqref{eq:ceff}}\!\!\!}}{=}\varOmega(t,0)+
 \frac{\ceff^{2}qA(t)}{\varOmega(t,0)}\kc+\order(\kc^{2})\text{.}
\end{align}

Comparing this result with Eq.~\eqref{eq:A(t)-gap-vic-Dirac-Omega} confirms
the analogy between the Dirac case and the semiconductor case up to the first
order in the conserved \mbox{(quasi)}mo\-men\-tum around $k=\kc=0$.

\subsection{Analogy in the entire Brillouin zone}

The analogy between the Hamiltonians can be extended to the whole
Brillouin zone, which means that each $\kc$ mode in $\HSC$ can be
mapped to a $k$ mode in $\HDirac$ with a suitable effective speed
of light and electron mass.
In the previous two subsections, we have derived that these effective quantities
are constant ($\kc$ independent) for long-wave\-length modes and that $k$
and $\kc$ have an interchangeable meaning for these modes.
However, this coincidence between $k$ and $\kc$ is not universal since we
can always confine the crystal momentum of a Bloch electron in the two-band
semiconductor to the first Brillouin zone (a consequence of restricting
ourselves to two bands), while each canonical wave vector in the Dirac case
represents a unique mode.
The distinction becomes important when we go beyond long-wave\-length modes,
which we will do in this subsection.

The question we aim to answer is as follows: Given a semiconductor band
struc\-ture---i.e., the functions
$\Delta\energy(\kc)$, $\Delta v(\kc)=\D\Delta\energy(\kc)/\D\kc$,
and $\kappa(\kc)$ are fixed---\allowbreak and a mode $\kc\in(-\pi/\LC,\pi/\LC]$,
can we then find effective constants $\meff(\kc)$ and $\ceff(\kc)$
and a wave vector $k=k(\kc)$ such that the eigenvalue $\Omega[t,k(\kc)]$
in the Dirac case {[}Eq.~\eqref{eq:A(t)-Dirac-Omega}{]} with these
$\kc$-de\-pen\-dent effective quantities equals the semiconductor analog
$\varOmega(t,\kc)$ in Eq.~\eqref{eq:A(t)-SC-Omega}?
We therefore want to solve the equation
\begin{multline}
\meff^{2}(\kc)\ceff^{4}(\kc)+\ceff^{2}(\kc)[k(\kc)+qA(t)]^{2}\\
{}=\left[\frac{\Delta\energy(\kc)+qA(t)\Delta v(\kc)}{2}\right]^{2}+
\left[\frac{qA(t)|\kappa(\kc)|}{m}\right]^{2}
\end{multline}
for an arbitrary potential $A(t)$, so we compare the coefficients
with respect to the powers of $A$.
This procedure yields three equations, which uniquely fix the three unknown
quantities
\begin{align}
\label{eq:ceff(K)}
\ceff(\kc) & =\sqrt{\frac{\Delta v^{2}(\kc)}{4}+
\frac{|\kappa(\kc)|^{2}}{m^{2}}}\text{,}
\\
\label{eq:meff(K)}
\meff(\kc) & =\frac{\Delta\energy(\kc)|\kappa(\kc)|}{2m\ceff^{3}(\kc)}\text{,}
\\
\label{eq:k(K)}
k(\kc) & =\frac{\Delta\energy(\kc)\Delta v(\kc)}{4\ceff^{2}(\kc)}\text{.}
\end{align}

These $\kc$-de\-pen\-dent effective quantities are of course compatible
with the results from the previous two subsections:
for $\kc=0$, we get $\ceff(0)=\kappa_{0}/m$,
$\meff(0)=\Eg m^{2}/(2\kappa_{0}^{2})$, and $k(0)=0$---ex\-act\-ly what we found
in Sec.~\ref{subsec:analogy-at-band-gap}.
Furthermore, we find that the first $\kc$ derivatives at $\kc=0$
are $\ceff^{(1)}(0)=\meff^{(1)}(0)=0$ and $k^{(1)}(0)=1$; that is,
for all long-wave\-length modes, the effective quantities are constant, and the
crystal momentum $\kc$ in the semiconductor has the same meaning as the momentum
$k$ in Dirac theory, which is basically the result of
Sec.~\ref{subsec:analogy-gap-vicinity}.

In the remainder of this paper, we will focus on modes with small conserved
\mbox{(quasi)}mo\-men\-ta again.
For brevity, we write $\ceff$ and $\meff$ without
a parameter again to denote the respective value at $\kc=0$.

Note that, in the gauge used here, $\kc$ is conserved exactly for purely
time-de\-pen\-dent fields $A(t)$.
This is somewhat different from other gauges where $\kc$ becomes effectively
time dependent $\kc\to \kc +qA(t)$, and thus the analogy between pair
creation and Landau--Zener tunneling during the temporal passage through an
avoided level crossing (at the gap $\kc=0$) becomes even more apparent.
In our representation (where $\kc$ is conserved), we may directly translate the
momentum spectra from QED calculations (e.g., for the dynamically assisted
Sauter--Schwinger effect~\cite{Orthaber+Hebenstreit+Alkofer2011,%
Fey+Schuetzhold2012}) to the semiconductor scenario via Eqs.~\eqref{eq:ceff(K)},
\eqref{eq:meff(K)}, and~\eqref{eq:k(K)}.
The only difference is that the range of $\kc$ is reduced to the Brillouin zone
in the semiconductor case, and the density of states is given per $\kc$ interval
(instead of $k$ for real QED), which introduces an additional factor of
$\D k/\D\kc$.

However, when comparing to experimental results, another important difference
must be taken into account:
the conserved wave numbers $\kc$ (and $k$) correspond to the canonical momenta,
which are generally different from the mechanical momenta.
The latter are not conserved, of course, because the electric field accelerates
the charged particles \emph{after} they have been created.
This acceleration then depends on the shape of the dispersion
relation, such that here the analogy to QED eventually breaks down.
Ergo, the analogy applies to the \emph{creation} of par\-ti\-cle--hole pairs
(for a given $\kc$), but not necessarily to their trajectory \emph{after}
they have been created.

\subsection{\label{subsec:Sauter-Schwinger-in-GaAs}%
Analog Sauter--Schwinger effect and dynamical assistance in gallium arsenide}

The fact that the Hamiltonians $\HDirac$ and $\HSC$ do coincide for
long-wave\-length modes (except for scales) allows us to infer
that we may directly transfer all findings regarding nonperturbative
(tunneling) pair creation from quantum electrodynamics to the semiconductor
model (at least to leading order).

\subsubsection{Constant electric field} 

Let us start with a constant electric field $\Estatic$ with $A(t)=\Estatic t$
as the simplest example.
In the Dirac case, this corresponds to the ordinary Sauter--Schwinger effect
with the associated critical electric field strength $\ES$
{[}see Eq.~\eqref{eq:ES}{]}.
In a semiconductor, interband tunneling due to a constant external
field is typically described via the Landau--Zener
model~\cite{Landau1932,Zener1932,Zener1934}---but due to the analogy with
quantum electrodynamics (QED), we may also use the QED terminology and
consequently define the analog critical field strength
\begin{equation}
\Ecrit=\frac{\ceff^{3}\meff^{2}}{q}=\frac{\sqrt{2\meff}\Eg^{3/2}}{4q}\text{.}
\label{eq:Ecrit}
\end{equation}
This expression, here simply derived from the analogy with QED, can be found in
many papers which study the behavior of semi\-con\-duc\-tors/in\-su\-la\-tors
in strong electric fields; see, e.g., Ref.~\cite{Kane1960}.

As an example for a semiconductor with a direct band gap at the Brillouin-zone
center (as assumed in Sec.~\ref{subsec:analogy-at-band-gap}), we
consider gallium arsenide (GaAs) here.
The band gap of GaAs measures about $\SC{\Eg}{GaAs}=1.5\,\mathrm{eV}$, and the
effective masses $\SC{\meffc}{GaAs}=0.063m$ and $\SC{\meffv}{GaAs}=0.076m$
(light holes; see Ref.~\cite{Sze+Kwok2006}) yield the value
$\SC{\meff}{GaAs}\approx0.069m$ according to Eq.~\eqref{eq:meff-of-meffs}.
The resulting critical field strength is thus
$\Ecrit[GaAs]\approx6.2\times10^{6}\,\mathrm{V/cm}$---a typical value for this
type of semiconductor according to, e.g., Ref.~\cite{Krieger+Iafrate1986}.
This value is roughly one order of magnitude larger than the
di\-elec\-tric-break\-down field strength of GaAs given in
Ref.~\cite{Sze+Kwok2006}.
This relation seems reasonable since interband tunneling starts below
$\Ecrit[GaAs]$ of course, but it is suppressed exponentially by the factor
$\exp(-\pi\Ecrit[GaAs]/\Estatic)$.
For $\Estatic\approx\Ecrit[GaAs]/10$, this factor measures $10^{-14}$.
We do not consider the (nonexponential) prefactor in the pair-cre\-a\-tion rate
here, but one can easily imagine that the exponential term suppresses any
realistic prefactor for much smaller values of $\Estatic$.

\subsubsection{Assisting temporal Sauter pulse} 

As a next example, we add a temporal Sauter pulse $\ESauter/\cosh^{2}(\omega t)$
to the constant background field $\Estatic$ and assume that the pulse
amplitude is much smaller than the static field; $\ESauter/\Estatic\ll1$.
The effect of the weak pulse on nonperturbative pair creation
has been studied in Ref.~\cite{Schuetzhold+Gies+Dunne2008}.
According to that paper, the pulse is negligible if its characteristic
frequency scale, $\omega$, is smaller than a certain critical value
$\omegacrit$, which depends on the background field strength but
(interestingly) not on the pulse amplitude.
This critical frequency scale is reached when the combined Keldysh parameter
\begin{equation}
\label{eq:combined-Keldysh-parameter}
\KP_{\omega} = \frac{m\omega}{q\Estatic} = \frac{\ES}{\Estatic} \frac{\omega}{m}
\end{equation}
takes on the value $\pi/2$.
Above this threshold, the so-called dynamically
assisted Sauter--Schwinger effect sets in, which means that
the pulse exponentially enhances the pure Sauter--Schwinger
pair-cre\-a\-tion rate induced by $\Estatic$.

Let us assume in our example that the background field is one order
of magnitude below the critical field strength.
In the Dirac case, that means $\Estatic=\ES/10$,
and we get a critical frequency scale in the hard X-ray spectrum:
$\omegacrit=80\,\mathrm{keV}$.
In our semiconductor example ($\Estatic=\Ecrit[GaAs]/10$), the result
$\omegacrit=0.12\,\mathrm{eV}$ lies in the infrared part of the spectrum.

\subsubsection{\label{subsubsec:const-backgr-and-ass-cos}
Assisting harmonic oscillation}

The last example profile consists of the constant background field $\Estatic$
again plus a harmonic oscillation $\Ecos\cos(\omega t)$.
Similar to the Sauter pulse, such a wave can increase the nonperturbative
pair-cre\-a\-tion rate exponentially as studied in Ref.~\cite{Linder+al2015}.
However, the critical value of the Keldysh
parameter~\eqref{eq:combined-Keldysh-parameter} for dynamical assistance
depends on the ratio $\Ecos/\Estatic$ for this pro\-file---or, if $\omega$ and
$\Estatic$ are fixed, we can inverse this relation to determine a critical
laser amplitude $\Ecoscrit$ as a function of $\Estatic$ and $\omega$.

In the worldline instanton picture, the effect of the additional oscillation
is that it lowers the instanton action $\InstAct$, which in turn increases the
pair-cre\-a\-tion rate since the rate is proportional to $\exp(-\InstAct)$.
(We ignore the nonexponential prefactor in the pair-cre\-a\-tion rate here;
however, it has been shown in Ref.~\cite{Schneider+Schuetzhold2016-2} that the
behavior of the exponent $\InstAct$ plays the crucial role in the dynamical
assistance mechanism.)
Let us (arbitrarily) define the threshold of dynamical assistance as a
configuration according to which the pair-cre\-a\-tion rate with the oscillation
{[}$\propto\exp(-\InstAct_\omega)${]} is 50\% larger than the rate
{[}$\propto\exp(-\InstAct_0)${]} in the constant background field $\Estatic$
alone.
We may derive from Eqs.~(52) and~(57) in Ref.~\cite{Linder+al2015} that this
condition gives
\begin{equation}
\frac{\E^{-\InstAct_\omega}}{\E^{-\InstAct_0}}
= \exp{\left[2\pi \frac{\ES}{\Estatic} \frac{I_1(\KP_\omega)}{\KP_\omega}
  \frac{\Ecos}{\Estatic} \right]}
\stackrel{!}{=} 1.5\text{,}
\end{equation}
where $I_1(x)$ denotes a modified Bessel function of the first kind.
Assuming that only the oscillation amplitude is variable, we solve this equation
for $\Ecos$ to find the critical amplitude
\begin{equation}
\Ecoscrit = \frac{\ln 1.5}{2\pi} \frac{\Estatic}{\ES}
            \frac{\KP_\omega}{I_1(\KP_\omega)} \Estatic \text{.}
\label{eq:Ecoscrit}
\end{equation}

Let us now transfer this QED result~\eqref{eq:Ecoscrit} to the semiconductor
analog and do some estimations regarding the experimental realization of
assisted tunneling pair creation in GaAs.
We assume a rather pure sample of GaAs placed in a background field
$\Estatic=\Ecrit[GaAs]/10$ again.
The harmonic oscillation is generated by a $\mathrm{CO_{2}}$ laser
with a wavelength of $10.6\,\mathrm{\mu m}$.
The corresponding photon energy, $0.117\,\mathrm{eV}$, measures less than
8\% of the band gap, so pair creation via multiphoton processes is strongly
suppressed.
The background field strength and the laser frequency together yield
the combined Keldysh parameter $\KP_{\omega}=1.56$.
While this value is fixed, we can easily vary the laser amplitude.
The critical amplitude~\eqref{eq:Ecoscrit} is then given by
$\Ecoscrit/\Estatic = 0.0097$ in this example, which corresponds to a laser-beam
intensity of $\Icrit = (\Ecoscrit)^2/2 = 47\,\mathrm{kW/cm^2}$.
References~\cite{Smith1972,Smith1973,Smith1974} (which consider only pulsed
radiation though) suggest that a GaAs sample of sufficient quality will probably
not be destroyed by this amount of incident pow\-er---the damage threshold for
$\mathrm{CO_2}$-laser pulses with a halfwidth of
$100\,\mathrm{ns}=10^{-7}\,\mathrm{s}$ given in Ref.~\cite{Smith1974} is of the
order of $10\,\mathrm{MW/cm^2}$, for example.
We will also show later (in Sec.~\ref{subsec:dyn-ass-spacetime-dep}) that the
threshold intensity is reduced significantly in a space-de\-pen\-dent static
background field of finite spatial extent.

The analog quantities given in this subsection including the values
for GaAs are summarized in Table~\ref{tab:A(t)-scale-subs}.


\begin{table}
\begin{tabular}{rcl}
\toprule
\textbf{Dirac theory} &  & \textbf{Two-band semiconductor}\tabularnewline
\midrule
electron mass $m$ & $\leftrightarrow$ & $\meff$, effective mass
{[}Eqs.~\eqref{eq:meff-of-kappa} and \eqref{eq:meff-of-meffs}{]}\tabularnewline
 &  & GaAs: $\meff\approx0.07m$\tabularnewline
\midrule
speed of light $\cvac$ & $\leftrightarrow$ & $\ceff=\sqrt{\Eg/(2\meff)}$,
effective speed\tabularnewline
 &  & GaAs: $\ceff\approx0.005\cvac$\tabularnewline
\midrule
mass gap $2m\cvac^{2}$ & $\leftrightarrow$ & $2\meff\ceff^{2}=\Eg$,
band gap\tabularnewline
${}\approx1\,\mathrm{MeV}$ & $\leftrightarrow$ & GaAs:
$\Eg\approx1.5\,\mathrm{eV}$\tabularnewline
\midrule
\multicolumn{3}{l}{\emph{Sauter--Schwinger effect:}}\tabularnewline
$\ES=m^{2}\cvac^{3}/q$ & $\leftrightarrow$ &
$\Ecrit=\sqrt{2\meff}\Eg^{3/2}/(4q)$\tabularnewline
${}\approx10^{18}\,\mathrm{V/m}$ & $\leftrightarrow$ & GaAs:
$\Ecrit\approx6\times10^{8}\,\mathrm{V/m}$\tabularnewline
\midrule
\multicolumn{3}{l}{\emph{Dynamically assisted Sauter--Schwinger
effect:}}\tabularnewline
$\omegacrit\approx80\,\mathrm{keV}$ & $\leftrightarrow$ & GaAs:
$\omegacrit\approx0.12\,\mathrm{eV}$\tabularnewline
\bottomrule
\end{tabular}%
\caption{%
\label{tab:A(t)-scale-subs}%
Comparison between the scales in the Dirac Hamiltonian and the analog
quantities in the semiconductor model.}
\end{table}

\boldmath
\section{Spacetime-dependent case $\Ef=\Ef(t,x)$}
\unboldmath
In this section, we generalize the semiconductor model
presented in the previous section to space\-time-de\-pen\-dent electric
fields and compare it to the corresponding Dirac Hamiltonian again.

\subsection{Hamiltonians}

Here, we choose a different gauge, $\Ef(t,x)=\partial_{x}\Phi(t,x)$,
with a vanishing vector potential $A$; that is, the field is described
by the space\-time-de\-pen\-dent scalar potential $\Phi$, which enters
the po\-si\-tion-space Hamiltonians $\HDirac$~\eqref{eq:A(t)-HDirac-pos-space}
and $\HSCfull$~\eqref{eq:A(t)-HSCfull} as an additional potential
term $-q\Phi$.
The mo\-men\-tum-space form of the Dirac Hamiltonian
thus contains the convolution of the spatial Fourier transform of
the scalar potential, $\Phim(t,k)$, and the mo\-men\-tum-space
field operator $\spi{\DFOm}$ {[}see Eq.~\eqref{eq:DFO-Fourier-trafo}
for the conventions we use{]}:
\begin{multline}
\label{eq:Phi(t,x)-HDirac-Fourier-space}
\HDirac(t)=\intop_{-\infty}^{\infty}\spi{\DFOm}^{\dagger}(t,k)
\begin{pmatrix}m & k\\k & -m\end{pmatrix}\spi{\DFOm}(t,k)\,\D k\\
{}-\frac{q}{\sqrt{2\pi}}\intop_{-\infty}^{\infty}\spi{\DFOm}^{\dagger}(t,k)
\intop_{-\infty}^{\infty}\Phim(t,k-k')\spi{\DFOm}(t,k')\,\D k'\,\D k\text{.}
\end{multline}

As in the time-de\-pen\-dent case, we want to bring this Hamiltonian
into a form in which the matrix in the upper line is diagonal.
This is accomplished by inserting the same transformed field operator
$\spi{\RDFO}$~\eqref{eq:A(t)-DFO-rotated} as in the previous section (but
with $A=0$ of course).
However, the very same transformation gives rise to a matrix $\PMD$ in the
lower ($\Phim$) part of the Hamiltonian.
This matrix reads as
\begin{align}
\PMD(k,k')={} & \frac{1}{\sqrt{1+d^{2}(k)}}\frac{1}{\sqrt{1+d^{2}(k')}}
\nonumber \\
 & {}\times\begin{pmatrix}1 & d(k)\\
-d(k) & 1 \end{pmatrix}\cdot\begin{pmatrix}1 & -d(k')\\ d(k') & 1 \end{pmatrix}
\end{align}
with the auxiliary function $d$ defined in Eq.~\eqref{eq:A(t)-Dirac-aux-d} but
without any time dependence here ($A=0$).
For the transformed Dirac Hamiltonian, we thus get
\begin{multline}
\HDirac(t)=\intop_{-\infty}^{\infty}\spi{\RDFO}^{\dagger}(t,k)
\begin{pmatrix}\Omega(k) & 0\\
0 & -\Omega(k) \end{pmatrix}\spi{\RDFO}(t,k)\,\D k\\
{}-\frac{q}{\sqrt{2\pi}}\intop_{\mathclap{-\infty}}^{\mathclap{\infty}}
\spi{\RDFO}^{\dagger}(t,k)\intop_{\mathclap{-\infty}}^{\mathclap{\infty}}
\Phim(t,k-k')\PMD(k,k')\spi{\RDFO}(t,k')\,\D k'\,\D k\text{,}
\label{eq:Phi(t,x)-HDirac-exact}
\end{multline}
again with the same (but time-in\-de\-pen\-dent) eigenvalues $\pm\Omega$
from Eq.~\eqref{eq:A(t)-Dirac-Omega}.

Let us now derive the semiconductor Hamiltonian in the space\-time-de\-pen\-dent
field.
We start with the full Hamiltonian~\eqref{eq:A(t)-HSCfull} again but with
$A=0$ and the additional potential term $-q\Phi$.
After the insertion of the two-band approximation~\eqref{eq:Two-band-field-op},
our semiconductor Hamiltonian reads as
\begin{multline}
\HSC(t)=\intop_{-\pi/\LC}^{\pi/\LC}\spi{\AOP}^{\dagger}(t,\kc)
\begin{pmatrix}\energy_{\cbi}(\kc) & 0\\
0 & \energy_{\vbi}(\kc)\end{pmatrix}\spi{\AOP}(t,\kc)\,\D\kc\\
-q\intop_{-\pi/\LC}^{\pi/\LC}\spi{\AOP}^{\dagger}(t,\kc)
\intop_{-\pi/\LC}^{\pi/\LC}\PMSC(t,\kc,\kc')\spi{\AOP}(t,\kc')\,\D\kc'\,\D\kc
\label{eq:Phi(t,x)-HSC-exact}
\end{multline}
with the matrix
\begin{equation}
\PMSC(t,\kc,\kc')=\begin{pmatrix}\braket{\cbi,\kc|\Phi|\cbi,\kc'} &
\braket{\cbi,\kc|\Phi|\vbi,\kc'}\\
\braket{\vbi,\kc|\Phi|\cbi,\kc'} & \braket{\vbi,\kc|\Phi|\vbi,\kc'}
\end{pmatrix}\!\text{.}
\label{eq:PMSC-exact}
\end{equation}
Further transformations of the operators $\AOP_{\covbi}$ are not
necessary in this case since the matrix in the upper line of $\HSC$
is already diagonal for the present gauge.

It is important to notice here that the diagonal elements $\energy_{\covbi}$
in $\HSC$ are generally not symmetric (for all $\kc$) as in the
Dirac case ($\pm\Omega$) in Eq.~\eqref{eq:Phi(t,x)-HDirac-exact}.
In the purely time-de\-pen\-dent field, we could make these diagonal elements
in $\HSC$ symmetric via a suitable gauge transformation
(see Sec.~\ref{subsec:A(t)-H-diagonalization}).
However, the same approach is not valid in a space\-time-de\-pen\-dent field
since the $\Phi$ part of the Hamiltonian couples particles with different values
of $\kc$ with each other, so $\kc$ is not a conserved quantity anymore, and thus
$\spi{\AOP}^{\dagger}(t,\kc)\spi{\AOP}(t,\kc)=1$ is \emph{not} valid in general
here for each $\kc$.
As we will see in the next subsection, this fact requires us to make an
additional assumption concerning the effective masses in the semiconductor in
order to draw the quantitative analogy to the Dirac Hamiltonian.

\boldmath
\subsection{\label{subsec:analogy-non-Phi-parts}Analogy between the
\texorpdfstring{$\Phi$}{Phi}-inde\-pen\-dent parts of the Hamiltonians}
\unboldmath
At this point, we can start to compare the upper lines
of the Hamiltonians $\HDirac$~\eqref{eq:Phi(t,x)-HDirac-exact}
and $\HSC$~\eqref{eq:Phi(t,x)-HSC-exact}, which do not depend on
the potential $\Phi$.
We focus on the vicinities of the band gaps at $k=\kc=0$ again.

Up to the lowest nonvanishing order of the small quantity $k/m$ near
the gap, the diagonal elements in the Dirac case are
\begin{equation}
\pm\Omega(k)=\pm m\cvac^{2}\pm\frac{k^{2}}{2m}+
\order{\left[\left(\frac{k}{m\cvac}\right)^{4}\right]}
\end{equation}
with $\cvac$ written explicitly.
According to our notion, the analogy to $\HSC$ is valid if $\pm\Omega$ coincides
with $\energy_{\covbi}$ {[}from Eq.~\eqref{eq:A(t)-SC-parabolic-bands}{]} up to
the quadratic order in $k$ or $\kc$ after substituting the physical scales $m$
and $\cvac$ with effective constants.
(As in Secs.~\ref{subsec:analogy-at-band-gap}
and~\ref{subsec:analogy-gap-vicinity} in the case of a purely time-de\-pen\-dent
$\Ef$ field, the physical roles of $k$ and $\kc$ are equivalent close to the
gaps.)

We find that the analogy works with the same effective constants,
$\ceff$~\eqref{eq:ceff} and $\meff$~\eqref{eq:meff-of-meffs},
as in the $A(t)$ case, but we have to assume in addition that the
effective electron mass in the conduction band, $\meffc$, equals
the effective hole mass $\meffv$ (in which case $\meff=\meffc=\meffv$).
Graphically, that means that the parabolic curvatures of the energy
curves $\energy_{\cbi}$ and $\energy_{\vbi}$ in Fig.~\subref{fig:disp-rel-SC}
must be identical at the gap.
From a practical point of view, this is an important constraint regarding
the simulation of nonperturbative vacuum pair production in
space\-time-de\-pen\-dent fields in semiconductors, which can only be met
approximately.
The effective masses in GaAs, $\SC{\meffc}{GaAs}=0.063m$ and
$\SC{\meffv}{GaAs}=0.076m$ (light holes), differ by about 20\%, for
ex\-ample---com\-pared to other common semiconductors with a direct band gap,
this is a quite good agreement (values taken from Ref.~\cite{Sze+Kwok2006}).

We will assume $\meff=\meffc=\meffv$ in the remainder of this section.

\boldmath
\subsection{\label{subsec:analogy-Phi-parts}Analogy between the
\texorpdfstring{$\Phi$}{Phi} parts for spatially slowly varying potentials}
\unboldmath
We still have to show that the analogy is also true for
the $\Phi$ parts {[}lower lines in Eqs.~\eqref{eq:Phi(t,x)-HDirac-exact}
and~\eqref{eq:Phi(t,x)-HSC-exact}{]} of the Hamiltonians in the
vicinity of the band gap.
We thus have to compare the matrix $\Phim(t,k-k')\PMD(k,k')/\sqrt{2\pi}$
in the Dirac case with $\PMSC(t,\kc,\kc')$ in the semiconductor case
since the other terms in the $\Phi$ parts are equivalent.
These matrices cannot be the same for arbitrary \mbox{(quasi)}mo\-men\-ta and
potentials $\Phi(t,x)$, so we have to make reasonable assumptions about these
quantities and then compare the matrices (approximately).

Let us start with the Dirac case.
As we can see in the Hamiltonian~\eqref{eq:Phi(t,x)-HDirac-exact},
the Fourier components of the potential $\Phi$ couple particle states
which differ by $k-k'$ in their wave vectors.
Since we want to concentrate on the parabolic vicinity of the band gap (the
range $|k|\ll m$) and electron transitions therein, we assume that the
potential only has nonvanishing Fourier components $\Phim(t,k)$ for small
wave vectors which satisfy $|k/m|\ll1$.
That is, the potential and thus the electric
field is slowly varying in space compared to the Compton wavelength
of an electron, and therefore an electron close to the gap cannot
be excited (directly) to a point far beyond the gap in $k$ space.

This assumption is also consistent with the fact that we are interested
in nonperturbative pair creation: for this reason, the electric field
should only incorporate photon energies far below the mass gap $2m$,
which correspond to wave numbers $|k|\ll2m$---lead\-ing basically
to the same assumption as above.

For a Dirac-sea electron with a $k$ near the gap ($|k/m|\ll1$),
which may be excited into another state with the small wave vector $k'$
($|k'/m|\ll1$) due to the potential, we may therefore Taylor expand
the matrix $\PMD(k,k')$ and neglect terms of second order in these small
wave vectors.
We get
\begin{multline}
\PMD(k,k')
= \begin{pmatrix}1 & 0\\0 & 1\end{pmatrix}
+ \frac{k-k'}{2m\cvac} \begin{pmatrix}0 & 1\\-1 & 0\end{pmatrix}
\vphantom{\order{\left[\frac{(k^{\prime})^2}{m^{2}\cvac^{2}}\right]}
+ \order{\left[\frac{k k^{\prime}}{m^{2}\cvac^{2}}\right]}} \\
\vphantom{\frac{k-k'}{2m\cvac} \begin{pmatrix}0 & 1\\-1 & 0\end{pmatrix}}
{} + \order{\left[\frac{k^{2}}{m^{2}\cvac^{2}}\right]}
+ \order{\left[\frac{(k^{\prime})^2}{m^{2}\cvac^{2}}\right]}
+ \order{\left[\frac{k k^{\prime}}{m^{2}\cvac^{2}}\right]}
\label{eq:Phi-part-Dirac-series}
\end{multline}
with the speed of light written explicitly.

In the semiconductor case, we have to approximate the matrix $\PMSC$
for slowly varying potentials.
These are potentials which only include wavelengths much greater than the
lattice constant $\LC$.
We therefore assume that its spatial Fourier transform, $\Phim(t,\kc)$,
vanishes except for $|\kc|\ll2\pi/\LC$.
In analogy to the Dirac case, this $\kc$-space region coincides with the
parabolic vicinity of the semiconductor band gap;
cf.\ Fig.~\subref{fig:disp-rel-SC}.

We think that this long-wave\-length assumption is practically always satisfied
in the context of nonperturbative elec\-tron--hole pair creation, which requires
the photon energies in the electric field to be much smaller than the band gap:
$\omega\ll\Eg$.
Let us do a simple estimate to show this:
Writing $\omega$ as $2\pi/(n\lambda)$, where $n$ is the refractive index in our
semiconductor (for the frequency under consideration), the condition
$\omega\ll\Eg$ becomes $\lambda\gg 2\pi/(n\Eg)$.
It is generally justified to assume that $\Eg$ is (much) smaller than the Fermi
energy $\EF=\pi^2/(2m\LC^2)$ in the empty lattice.
Inserting this relation into the above inequality lets us conclude that
$\lambda\gg 2\pi/(n\EF)$, which can also be written as
$\lambda\gg (8/n)(\LC/\lambdaC)\LC$, where
$\lambdaC\approx 10^{-12}\,\mathrm{m}$ is the Compton wavelength of the
electron.
For typical semiconductors, $\LC/\lambdaC$ is much greater than $1$, while $8/n$
is of order $1$.
Hence, $\lambda\gg\LC$ should be reasonable to assume provided $\omega\ll\Eg$
for all photons in the external field.

Since we are especially interested in GaAs here, let us consider this case
in particular:
The assumption $\omega\ll\SC{\Eg}{GaAs}=1.5\,\mathrm{eV}$ corresponds to vacuum
wavelengths much greater than $816\,\mathrm{nm}$.
The refractive index of GaAs around the band gap measures about $3.7$
(see Ref.~\cite{RefractiveIndex.info} and cf., e.g., Ref.~\cite{Sze+Kwok2006}),
so the wavelengths within the medium must be much greater than approximately
$220\,\mathrm{nm}$---a length scale which is very large compared to the
lattice constant $0.565\,\mathrm{nm}$ of GaAs.
The assumption of a slowly varying potential in the semiconductor case is
thus not problematic in the context of nonperturbative pair creation in GaAs, 
and, as argued above, this statement presumably also holds in most other
semiconductors.

This assumption together with the fact that we consider quasimomenta
obeying $|\kc|\ll2\pi/\LC$ and $|\kc'|\ll2\pi/\LC$ lets us derive
the (still exact) expression
\begin{multline}
\PMSC(t,\kc,\kc')=\frac{1}{\sqrt{2\pi}}\Phim(t,\kc-\kc')\\
{}\times\begin{pmatrix}
\braketcell{\cbi,\kc|\cbi,\kc'} & \braketcell{\cbi,\kc|\vbi,\kc'}\\
\braketcell{\vbi,\kc|\cbi,\kc'} & \braketcell{\vbi,\kc|\vbi,\kc'}\end{pmatrix}
\label{eq:PMSC-for-slow-Phi}
\end{multline}
for the matrix in Eq.~\eqref{eq:PMSC-exact}; see
Appendix~\ref{app:PMSC-for-slow-Phi} for the calculation.

Since we are close to the band gap, we may expand the Bloch factors
which appear in the $\braketcell{\dots}$ products {[}defined in
Eq.~\eqref{eq:Bloch-factor-orthonormalization}{]} around $\kc=0$ up to the first
order in $\kc$ or $\kc'$ using $k\cdot p$ perturbation theory and the two-band
approximation again (cf.\ Appendix~\ref{app:kappa-Taylor-exp}).
Inserting these expansions from Eq.~\eqref{eq:u-kp-expansion} and
also using the Bloch-fac\-tor orthonormality
relation~\eqref{eq:Bloch-factor-orthonormalization} yields
\begin{multline}
\PMSC(t,\kc,\kc')
=
\frac{\Phim(t,\kc-\kc')}{\sqrt{2\pi}}
\left\{
\begin{pmatrix}1 & 0\\0 & 1\end{pmatrix} + \frac{\kappa_{0}\,(\kc-\kc')}{m\Eg}
\right.\\
\left.
\vphantom{\frac{\kappa_{0}\,(\kc-\kc')}{m\Eg}}
{} \times \begin{pmatrix}0 & 1\\-1 & 0\end{pmatrix}
+ \order[\kc^{2}] + \order[(\kc')^{2}] + \order[\kc \kc']
\right\}
\text{.}
\label{eq:Phi-part-SC-step}
\end{multline}

Let us now identify the correct effective scales: We consider the expression
$2\meff\ceff$.
According to Eq.~\eqref{eq:ceff}, this quantity is equal to $\sqrt{2\meff\Eg}$,
which in turn becomes $m\Eg/\kappa_{0}$ by means of
Eq.~\eqref{eq:meff-of-kappa}.
We can thus write $\PMSC$ as
\begin{multline}
\PMSC(t,\kc,\kc')
= \frac{\Phim(t,\kc-\kc')}{\sqrt{2\pi}}
\left\{
\begin{pmatrix}1 & 0\\0 & 1\end{pmatrix}
+ \frac{\kc-\kc'}{2\meff\ceff}
\right. \\
\left.
\vphantom{\frac{\kc-\kc'}{2\meff\ceff}}
\times \begin{pmatrix}0 & 1\\-1 & 0\end{pmatrix}
+ \order[\kc^{2}]
+ \order[(\kc')^{2}]
+ \order[\kc \kc']
\right\} \text{.}
\end{multline}
Comparing this equation to Eq.~\eqref{eq:Phi-part-Dirac-series}
shows that the $\Phi$ parts of the Hamiltonians are equivalent close
to the band gaps as well, with the same scale substitutions as before.
Hence, we have derived the analogy between $\HDirac$ and $\HSC$
also in the space\-time-de\-pen\-dent case.

\subsection{\label{subsec:dyn-ass-spacetime-dep}Dynamically assisted
Sauter--Schwinger effect in the space\-time-de\-pen\-dent case}

We close this section by considering an experimentally oriented setup,
which is a space\-time-de\-pen\-dent version of the dynamically assisted
Sauter--Schwinger effect in a semiconductor analog.

\subsubsection{Assisting temporal Sauter pulse} 

\begin{figure}
\subfloat[%
\label{fig:spatial-Sauter-supercrit}%
Large potential step $q\Delta\Phi\gg\chi$.]{\includegraphics{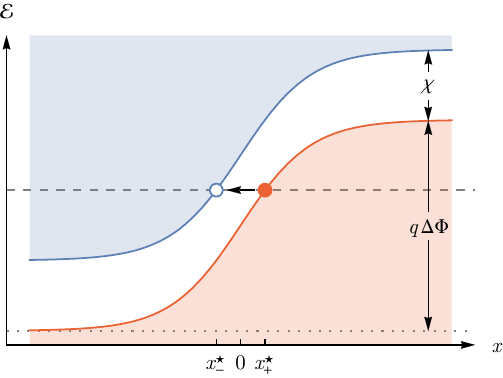}}

\subfloat[%
\label{fig:spatial-Sauter-crit}%
Slightly above threshold $q\Delta\Phi=\chi$.]{\includegraphics{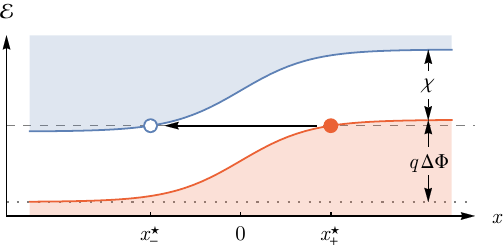}}

\subfloat[%
\label{fig:spatial-Sauter-subcrit}%
No tunneling for $q\Delta\Phi<\chi$.]{\includegraphics{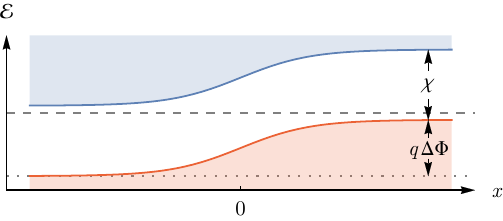}}

\caption{\label{fig:spatial-Sauter}%
Two energy bands separated by a gap $\chi$ are bent by a localized,
space-de\-pen\-dent electric field centered at $x=0$
(schematically; $\tanh$ profiles).
The solid curves are the lower edge of the upper energy
band and the upper edge of the lower band, respectively.
We have $\chi=2m$ in QED and $\chi=\Eg$ in the semiconductor analog.
(\protect\subref*{fig:spatial-Sauter-supercrit})~%
For field profiles which give rise to a large potential difference
$q\Delta\Phi\gg\chi$, there are many different states between which tunneling
is possible (e.g., along the dashed line).
(\protect\subref*{fig:spatial-Sauter-crit})~%
For $q\Delta\Phi\searrow\chi$, the number of possible tunneling transitions
approaches zero and the spatial turning points $\xT_{\pm}$ diverge.
In this $\Delta\Phi$ range, the tunneling rate can significantly be increased
via additional electric low-fre\-quen\-cy pulses according to
Ref.~\cite{Schneider+Schuetzhold2016-1}.
(\protect\subref*{fig:spatial-Sauter-subcrit})~%
If the energy step $q\Delta\Phi$ is smaller than the gap $\chi$,
the bands are separated energetically as indicated by the dashed
con\-stant-en\-er\-gy line, so tunneling becomes impossible.}
\end{figure}

A space\-time-de\-pen\-dent QED scenario has been studied analytically
in Ref.~\cite{Schneider+Schuetzhold2016-1} via the worldline instanton method.
In this reference, the superposition of a spatial and a temporal Sauter pulse
is considered:
\begin{equation}
\Ef(t,x) = \frac{\Ef_1}{\cosh^{2}(kx)} + \frac{\Ef_2}{\cosh^{2}(\omega t)}
\text{.}
\end{equation}
If the spatial pulse is very broad {[}quasi-homo\-ge\-neous, cf.\ the linear,
middle part in the band diagram in
Fig.~\subref{fig:spatial-Sauter-supercrit}{]}, the ``ordinary'' dynamically
assisted Sauter--Schwinger effect~\cite{Schuetzhold+Gies+Dunne2008} known from
the purely time-de\-pen\-dent case in
Sec.~\ref{subsec:Sauter-Schwinger-in-GaAs} is recovered.
This effect starts at
\begin{equation}
\KPcrit_{\omega} = \frac{m \omegacrit}{q\Ef_1} = \frac{\pi}{2}\text{.}
\end{equation}
The spatial turning points (between which tunneling happens) read as
$\xT_{\pm}=\pm m/(q\Ef_{1})$ in this case.

This situation changes when we narrow the spatial pulse $\Ef_{1}/\cosh^{2}(kx)$
by increasing its $k$, while $\Ef_{1}$ is kept constant and subcritical
here ($\Ef_{1}\ll\ES$).
The total electrostatic energy the pulse provides reads as
$q\Delta\Phi=2q\Ef_{1}/k$.
When this energy approaches the mass gap from above, $q\Delta\Phi\searrow2m$,
due to an increasing $k$, we get a band diagram like in
Fig.~\subref{fig:spatial-Sauter-crit}.
The spatial turning points grow according to
$\xT_{\pm}\sim\pm|\ln(q\Delta\Phi-2m)|$ in this limit, so the tunneling rate
due to the spatial pulse alone is low then.
These turning points are also the positions in Euclidean spacetime where the
corresponding instanton trajectory (we are just referring to the spatial Sauter
pulse at the moment) crosses the spatial axis ($x$).
The positions $\pm\ti_0$ where this instanton trajectory intersects the $\ti$
axis (imaginary time, $\ti=\I t$) are given by
\begin{equation}
\ti_0 = \frac{m}{q\Ef_1} \frac{\arcsin \KP_k}{\KP_k \sqrt{1-\KP_{k}^{2}}}
\label{eq:tau-turning_spat-Sauter}
\end{equation}
according to Ref.~\cite{Dunne+Schubert2005}, where
\begin{equation}
\KP_k = \frac{mk}{q\Ef_1}
\label{eq:spatial-KP}
\end{equation}
is the Keldysh parameter of the spatial Sauter pulse.
Hence, the positions $\pm\ti_0$ diverge like $1/\sqrt{q\Delta\Phi-2m}$ in the
limit $q\Delta\Phi\searrow2m$, which is equivalent to $\KP_k \nearrow 1$;
see Refs.~\cite{Schneider+Schuetzhold2016-1}.

The effect of the additional temporal Sauter pulse $\propto\cosh^{-2}(\omega t)$
is, in the instanton picture, that it gives rise to ``walls'' at
$\pm\ti_{\mathrm{sing}}=\pm\pi/(2\omega)$, which ``reflect'' the instanton
trajectory when touched.
The value of $\omega$ for which the unperturbed (no temporal pulse) instanton
trajectory just touches these ``walls'' is precisely $\omegacrit$,
the onset frequency scale for dynamical assistance.
The in\-stan\-ton-tra\-jec\-to\-ry scalings explained above let us conclude that
$\omegacrit\sim\sqrt{q\Delta\Phi-2m}$ in the limit $q\Delta\Phi\searrow2m$;
cf.\ Ref.~\cite{Schneider+Schuetzhold2016-1}.
Hence, if $q\Delta\Phi$ is only slightly larger than $2m$
{[}Fig.~\subref{fig:spatial-Sauter-crit}{]}, even low-fre\-quen\-cy pulses
should lead to an exponential enhancement of nonperturbative (tunneling) pair
cre\-a\-tion via the dynamically assisted Sauter--Schwinger effect.
Since the space dependence of such pulses is slow, their
purely time-de\-pen\-dent treatment should be valid.


Now let us transfer this situation to the semiconductor analog.
A localized, time-in\-de\-pen\-dent $\Ef$ field within a semiconductor
gives rise to the same schematic band diagrams depicted in
Fig.~\ref{fig:spatial-Sauter}.
For example, the band bending may be due to a suitable doping profile plus an
additional external bias if required.
The exact form of the bands will not be that of a hyperbolic tangent in general
as in Fig.~\ref{fig:spatial-Sauter}, which corresponds to a spatial
Sauter-pulse $\Ef$ field $\propto\cosh^{2}(kx)$.
However, we assume that the spatial $\Ef$ field within the semiconductor
does also decay exponentially for large $|x|$, just like a spatial
Sauter pulse does---but we do not prescribe an exact pulse shape near the
field maximum (the region around $x=0$ in Fig.~\ref{fig:spatial-Sauter}).
Note that this assumption is not compatible with the conventional depletion
approximation (see, e.g., Ref.~\cite{Sze+Kwok2006}) according to which the
density of ionized dopants is piecewise constant, which leads to parabolic
potential curves within these ionized regions and constant potential values
outside.
But, the idea of sharp transitions between ionized and unionized regions is
generally considered unrealistic, and one expects ``smeared'' transitions
instead (see Ref.~\cite{Sze+Kwok2006}).
We think that it is physically reasonable to assume exponential ``tails'' at the
edges of such transitions, which, together with the Boltzmann statistics of the
free charge carriers, should lead to a built-in field approaching zero
exponentially (far away from $x=0$).
Even if thermal effects are negligible (low temperatures), we nevertheless still
expect the built-in field to decay exponentially due to quantum effects:
if we think of the ionized, spatially fixed dopants on either side of the
junction as creating an effective, finite potential well for the respective
majority carriers, the wave functions of these carriers will leak into the
forbidden region (which begins somewhere on the other side of the
junc\-tion)---\allowbreak an effect which is in accordance with the exponential
decay of the built-in field.

Assuming that the time-in\-de\-pen\-dent (built-in) field within the
semiconductor decays exponentially, we conclude that the spatial turning
points $\xT_{\pm}$ scale like $\ln(q\Delta\Phi-\Eg)$ in the critical
limit $q\Delta\Phi\searrow\Eg$ {[}Fig.~\subref{fig:spatial-Sauter-crit}{]},
just like in the QED case above.
Now, let us imagine the unperturbed (no additional temporal Sauter pulse)
instanton trajectory in this limit: $\xT_{\pm}$ will be large, so the
instanton trajectory will be a huge closed loop over the $x$ range
$[\xT_{-},\xT_{+}]$.
Except near $x=0$, where we do not know the exact pulse shape of the spatial
field in the semiconductor, the instanton trajectory is the same
as that of a spatial Sauter pulse (QED case above) because the $\Ef$
fields in both cases decay exponentially; this functional form is sufficient
to fix the shape of the instanton trajectory.
The imag\-i\-nary-time ($\ti$) positions where the instanton trajectory
crosses the $\ti$ axis will thus also diverge like $1/\sqrt{q\Delta\Phi-\Eg}$
in the limit $q\Delta\Phi\searrow\Eg$.
This is because the exponential tails of the field let the instanton
trajectory grow so large in this limit that the details close to the
maximum field (around $x=0$) are not important for the scaling anymore.
Consequently, we expect the same scaling,
$\omegacrit\sim\sqrt{q\Delta\Phi-\Eg}$, as in the QED
case~\cite{Schneider+Schuetzhold2016-1} to be exhibited by an analog of the
dynamically assisted Sauter--Schwinger effect in a semiconductor with a
localized, time-in\-de\-pen\-dent inner field in the limit
$q\Delta\Phi\searrow\Eg$ as well.

Note that this scaling law solely depends on the way the electric field
approaches zero asymptotically (here: exponentially).
See Refs.~\cite{Gies+Torgrimsson2016,Gies+Torgrimsson2017}
for more information on universal pair-cre\-a\-tion phenomena in
the no-tun\-nel\-ing limit.

\subsubsection{Assisting harmonic oscillation} 

Another way to assist tunneling dynamically in this space\-time-de\-pen\-dent
scenario is via a harmonic oscillation instead of a temporal Sauter pulse.
This profile,
\begin{equation}
\Ef(t,x) = \frac{\Ef_1}{\cosh^{2}(kx)} + \Ef_2 \cos(\omega t)\text{,}
\end{equation}
is more appropriate to describe experiments in which pair creation
is assisted via laser beams, for example.
The purely time-de\-pen\-dent version of this profile (homogeneous background
field instead of a spatial Sauter pulse) has been studied in
Ref.~\cite{Linder+al2015}, also via the worldline instanton method.
In contrast to the temporal Sauter pulse, the oscillation does not give rise
to ``walls'' (singularities) parallel to the $x$ axis in Euclidean spacetime
because $\cos(\omega t)=\cosh(\omega\ti)$ is well behaved for all imaginary
times $\ti=\I t$.
Hence, the onset of dynamical assistance by the oscillation is not as sharply
defined as in the Sauter-pulse case.
We have formulated the threshold condition~\eqref{eq:Ecoscrit} for the
oscillation amplitude $\Ef_2$ in the case of a homogeneous background field
($k=0$ limit) in Sec.~\ref{subsubsec:const-backgr-and-ass-cos}
(with $\Estatic\to\Ef_1$ and $\Ecos\to\Ef_2$ here).
Let us now estimate how this critical oscillation amplitude $\Etwocrit$
changes when the background field becomes a spatial Sauter pulse ($k>0$), while
the maximum background field strength $\Ef_1$ and the oscillation frequency
$\omega$ remain fixed.

We consider the instanton trajectory of the spatial Sauter pulse again (see
Ref.~\cite{Dunne+Schubert2005}).
This closed loop in Euclidean spacetime has its largest extent
{[}from $-\ti_0$ to $+\ti_0$ with $\ti_0$ from
Eq.~\eqref{eq:tau-turning_spat-Sauter}{]} in the imag\-i\-nary-time direction on
the $\ti$ axis ($x=0$), where the field strength of the spatial Sauter pulse
measures $\Ef_1$---and that of the oscillation would be
$\Ef_2 \cosh(\omega\ti_0)$.
We assume that this instanton trajectory will be noticeably deformed (dynamical
assistance) by the additional oscillation if the amplitude $\Ef_2$ is large
enough such that the term $\Ef_{2}\cosh(\omega\ti_0)$ has a magnitude comparable
to $\Ef_1$.
Equation~\eqref{eq:Ecoscrit} can be understood as defining a certain ``threshold
ratio'' between these two terms for the special case of a homogeneous background
field {[}$k=0$, in which case $\ti_0 = m/(q\Ef_1)${]}:
\begin{equation}
\frac{\Etwocrit(k=0)}{\Ef_1} \cosh[\underbrace{\omega\ti_0(k=0)}_{\KP_\omega}]
\stackrel{!}{=} \text{const.}
\end{equation}
When we now increase $k$ (i.e., decrease the pulse width), $\ti_0(k)$ grows
according to Eq.~\eqref{eq:tau-turning_spat-Sauter}.
As a simple estimate, we determine the critical amplitude $\Etwocrit(k)$ for
this nonzero $k$ by demanding that the constant on the right-hand side of the
above equation remains invariant.
Hence, $\Etwocrit(k>0)$ must be smaller than $\Etwocrit(k=0)$ to compensate the
increase of $\cosh[\omega\ti_0(k>0)]$.
By considering the ratio between both critical amplitudes, we can eliminate the
constant and find
\begin{equation}
\frac{\Etwocrit(k)}{\Etwocrit(k=0)}
= \frac{\cosh \KP_\omega}{\cosh[\omega\ti_0(k)]} \text{.}
\label{eq:Ecrit-of-k-cos}
\end{equation}
Note that this way to derive $\Etwocrit(k)$ is not guaranteed to preserve the
property that we have originally imposed to find the critical amplitude in
the ho\-mo\-ge\-neous-field case (the oscillation enhances the pair-cre\-a\-tion
yield by 50\%; see Sec.~\ref{subsubsec:const-backgr-and-ass-cos})---rath\-er, we
have presented a simple way to estimate how $\Etwocrit(k)$ changes when
increasing $k$ from zero, and our main intention here is to show that the
critical amplitude decreases when the spatial extent of the static background
field gets smaller.

By squaring Eq.~\eqref{eq:Ecrit-of-k-cos} (and inserting $\ti_0$), we finally
find an expression for the critical (laser-beam) intensity as a function of the
inverse Sauter-pulse length scale $k$:
\begin{equation}
\frac{\Icrit(k)}{\Icrit(k=0)}
= \frac{\cosh^2 \KP_\omega}{\cosh^2{\left[\KP_\omega \arcsin(\KP_k)/
{\left(\KP_k \sqrt{1-\KP_{k}^{2}}\right)}\right]}} \text{,}
\label{eq:Icrit-of-k}
\end{equation}
where the threshold for a constant background field, $\Icrit(k=0)$, can be
calculated via Eq.~\eqref{eq:Ecoscrit}.
Note that $\Icrit(k)$ decreases for increasing $k$ until the critical amplitude
becomes zero at a certain $k$ value with $\KP_k = 1$.
This is precisely the $k$ value at which tunneling due to the spatial Sauter
pulse alone vanishes {[}cf.\ Fig.~\subref{fig:spatial-Sauter-crit}{]}, so the
concept of assisted tunneling breaks down there.
Hence, by decreasing the width of the static background field appropriately, we
can make the threshold intensity for dynamical assistance via the oscillation
arbitrarily small in principle---how\-ev\-er, in order to really verify this
effect under controlled conditions in the laboratory, the tunneling currents
(assisted and non-as\-sisted) should not become too tiny, so that they remain
measurable.
This requirement poses a practical limit on how narrow the spatial Sauter pulse
(built-in field) may become.

\begin{figure}
\includegraphics{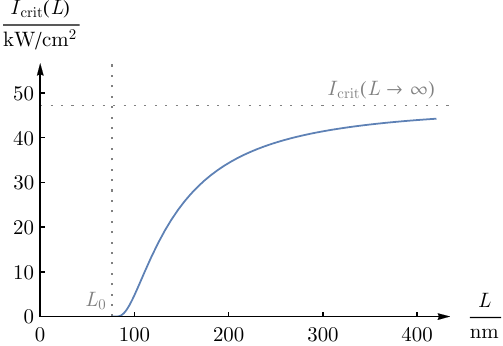}
\caption{\label{fig:I-crit_of_L}%
Threshold $\mathrm{CO_2}$-laser-beam intensity~\protect\eqref{eq:Icrit-of-k} for
dynamical assistance of tunneling as a function of the width $L=2\pi/k$ of the
static Sauter pulse $\Ef_1 / \cosh^2 (kx)$ in GaAs.
The parameter values in this plot are
$\Ef_1 = \Ecrit[GaAs]/10 \approx 60\,\mathrm{MV/m}$,
$\omega = 0.117\,\mathrm{eV}$ (so $\KP_\omega = 1.56$), and
$\Icrit(L \to \infty) = \Icrit(k=0) = 47\,\mathrm{kW/cm^2}$ (see
Sec.~\protect\ref{subsubsec:const-backgr-and-ass-cos}).
Tunneling vanishes in the limit $\KP_k \nearrow 1$
{[}cf.\ Fig.~\protect\subref{fig:spatial-Sauter-crit}{]}, which corresponds to
$L \searrow L_0 = 76\,\mathrm{nm}$ here.}
\end{figure}

Let us exemplify the result of Eq.~\eqref{eq:Icrit-of-k} for a semiconductor
analog by reconsidering the experimental setup from
Sec.~\ref{subsubsec:const-backgr-and-ass-cos} (time-de\-pen\-dent case; i.e.,
homogeneous fields only):
we said there that tunneling pair creation in GaAs induced by a constant
background field $\Ef_1 = \Ecrit[GaAs]/10 \approx 60\,\mathrm{MV/m}$ will
significantly be assisted by a $\mathrm{CO_2}$-laser wave $\Ef_2 \cos(\omega t)$
(with $\omega = 0.117\,\mathrm{eV}$ fixed, so $\KP_\omega = 1.56$) if the beam
intensity is about $\Icrit(k=0) = 47\,\mathrm{kW/cm^2}$.
If we replace the constant background field with a spatial Sauter pulse
$\Ef_1 / \cosh^2 (kx)$ with an associated length scale $L=2\pi/k$,
Eq.~\eqref{eq:Icrit-of-k} gives us the $L$-de\-pen\-dent critical laser
intensity plotted in Fig.~\ref{fig:I-crit_of_L}.


We emphasize that the dynamical assistance mechanisms from
Refs.~\cite{Schuetzhold+Gies+Dunne2008,Linder+al2015,%
Schneider+Schuetzhold2016-1}
considered here are fully nonperturbative effects, which are based
on a clas\-si\-cal-field description of the external fields.
So, even though we assume the assisting temporal Sauter pulse and the
time-de\-pen\-dent oscillation to be weak in amplitude
($\Ef_{2}\ll\Ef_{1}\ll\ES$), they still must incorporate a large number of
photons (high intensity) as to allow for the clas\-si\-cal field picture.
The dynamically assisted Sauter--Schwinger effect in semiconductors should thus
not be confused with the Franz--Keldysh effect~\cite{Franz1958,Keldysh1958-2}
(see also Refs.~\cite{Weiler+Zawadzki+Lax1967}), which is related to a shift in
the pho\-ton-ab\-sorp\-tion edge.
The QED analog of this effect was considered in Refs.~%
\cite{Dunne+Gies+Schuetzhold2009,Monin+Voloshin2010-1,Monin+Voloshin2010-2}.

\section{Generalization to electromagnetic fields in 2+1 dimensions}

In this section, we briefly discuss the feasibility to generalize
the analogy between Bloch electrons and holes in semiconductors and
Dirac's theory to 2+1 spacetime dimensions, including known results.

The step from one to two spatial dimensions is interesting because
it also allows for external magnetic fields, not just electric fields
as in the one-di\-men\-sion\-al case.
The Dirac field operator $\spi{\DFO}$ still has two components in two dimensions
since there is a third Pauli matrix ($\sigma_{x}$) for the additional required
gamma matrix $\gamma^{2}$.
This absence of spin simplifies the calculations and is typically irrelevant
in the context of tunneling pair
creation~\cite{Weiler+Zawadzki+Lax1967,Dunne+Schubert2005}.
In two-di\-men\-sion\-al space, the magnetic field is scalar and acts like the
$B_{z}$ component for charge carriers confined to the $(x,y)$ plane in three
dimensions.
It is given by the components of the vector potential $\vec{A}(t,x,y)$
via $B=-\partial_{x}A_{y}+\partial_{y}A_{x}$.

Graphene (see Refs.~\cite{Novoselov+al2005,CastroNeto+al2009}) is a well-known
example for a two-di\-men\-sion\-al system which mimics relativistic electron
motion near the points where the conduction band touches the valence band in
the Brillouin zone (Dirac cones); see also Ref.~\cite{Zawadzki2013}.
However, the associated effective electron rest mass is zero, so the analog of
the Schwinger limit $\ES\propto m^2$ vanishes in graphene, and thus there is no
characteristic exponential suppression of the Sauter--Schwinger effect; see
Refs.~\cite{Allor+Cohen+McGady2008,Dora+Moessner2010}.
But by generating an offset (symmetry breaking) between the two triangular
carbon lattices, which in combination make up the honeycomb structure
of graphene, it is possible to separate both energy bands by a finite energy
gap.
The Dirac cones of this so-called semiconducting graphene become shaped
like paraboloids near the gaps, which corresponds to a nonvanishing
effective rest mass.
Semiconducting graphene has already been produced successfully in the laboratory
via epitaxial growth as reported in Ref.~\cite{Nevius+al2015}, and it has been
studied in Ref.~\cite{Akal+al2016} as an analog for elec\-tron--pos\-i\-tron
pair creation in constant and oscillating (in time) electric fields.

One possible problem with analogs of Dirac's theory in multiple space
dimensions is that the vacuum is isotropic, so $m$ and $\cvac$ are
scalar quantities, while material properties of semiconductors, for
example, can depend on direction (effective mass tensor,
di\-rec\-tion-de\-pen\-dent effective speed of light, etc.).
Since these anisotropies have no counterpart in Dirac theory, we focus on
materials which behave isotropically around the band gap (scalar effective
quantities) or at least whose anisotropies do not interfere for the
electromagnetic field profile under consideration.

A simple profile which is interesting to study in 2+1 dimensions consists
of perpendicular electric ($x$ direction) and magnetic fields, both
constant.
In Dirac theory, the magnetic field decreases the pair-cre\-a\-tion
rate induced by the $\Ef$ field because we can always Lorentz-trans\-form
to a frame according to which the magnetic field is zero and the
pair-cre\-at\-ing electric field measures $\Ef^{2}-B^{2}$; see, e.g.,
Ref.~\cite{Schwinger1951,Landau+Lifshitz1971}.
That means that Sauter--Schwinger pair creation vanishes
completely for strong enough magnetic fields ($B=\Ef/\cvac$ or higher
in SI units).
In Ref.~\cite{Aronov+Pikus1967-1}, the authors state that for the same reason
the equivalent effect also happens in a two-band semiconductor, but again
with the effective scales $m\to\meff$ and $\cvac\to\ceff$
{[}Eqs.~\eqref{eq:ceff} and~\eqref{eq:meff-of-meffs}{]}.
(As in Sec.~\ref{subsec:analogy-non-Phi-parts}, we have to assume
$\meffc=\meffv=\meff$ here.)
Their reasoning is that the electrons in the semiconductor obey an effective
Dirac equation (near the band gap) since this type of equation models a simple
two-band system.
The validity of a Dirac-type equation implies the existence of an
analog Lorentz transformation (with $\cvac\to\ceff$), which is then
used to show that tunneling vanishes for $B=\Ef/\ceff$ in the semiconductor.
More detailed explanations of this Dirac-type two-band model are given
in Refs.~\cite{Zawadzki+Lax1966,Weiler+Zawadzki+Lax1967}, which
also study the crossed-field profile, and in
Refs.~\cite{Zawadzki+Rusin2011,Zawadzki2013}.

We can understand the reduction of Landau--Zener tunneling
in a semiconductor due to a perpendicular $B$ field as well by starting
with the same approach as in the previous sections, which deal with the
QED--semi\-con\-duc\-tor analogy in 1+1 dimensions.
That is, we begin with the Schr\"{o}dinger Hamiltonian~\eqref{eq:A(t)-HSCfull}
again but for 2+1 dimensions and with the vector potential
$\vec{A}(x)=-Bx\vec{e}_{y}$ and the additional scalar potential $\Phi(x)=\Ef x$
(crossed constant fields).
We then insert the 2+1-di\-men\-sion\-al version of the two-band
approximation~\eqref{eq:Two-band-field-op}.
The resulting two-band
Hamiltonian contains first- and sec\-ond-or\-der derivatives with
respect to the component $\kc_{x}$ of the crystal momentum, which
arise from the Bloch-basis representations of $x$ and $x^{2}$ (see,
e.g., Refs.~\cite{Adams1952,Adams1953,Gu+Kwong+Binder2013} for the
calculation of these matrix elements).
As a simple, semiclassical approach,
we then consider just the center of the Brillouin zone $\vec{\kc}=0$
(where we, again, assume the direct band gap to be located) and derive
the corresponding $x$-de\-pen\-dent band energies from the Hamiltonian
($\I\partial_{\kc_{x}}\to x$).
What we find is an expression which looks similar to the relativistic
counterpart
\begin{equation}
\energy_{\pm}(x) = -q\Ef x \pm \sqrt{m^{2}\cvac^{4} + (\cvac qBx)^2}
\end{equation}
(for the same crossed-field profile and $\vec{k}=0$) but with the
known effective constants $\cvac\to\ceff$ and $m\to\meff$, plus
additional terms under the square root.
However, these additional terms can be neglected for typical values
$\meff/m\approx10^{-2}\text{--}10^{-1}$,
$\ceff/\cvac\approx10^{-3}\text{--}10^{-2}$ (see the data for GaAs in
Table~\ref{tab:A(t)-scale-subs}, for example), a not too strong
tun\-nel\-ing-in\-duc\-ing electric field
$\Ef\approx10^{-1}\Ecrit\approx10^{7}\,\mathrm{V/m}$, a perpendicular magnetic
field in the range $B\lesssim\Ef/\ceff\approx10\,\mathrm{T}$, and $x$ values of
the order of the unperturbed (by the $B$ field) tunneling length $\Eg/(q\Ef)$.
The $\energy_{\pm}(x)$ graphs in the semiconductor thus look like the
relativistic version, which was also found in
Ref.~\cite{Weiler+Zawadzki+Lax1967}.

We emphasize that the reduction of the tunneling current in perpendicular
$B$ fields has just been explained by referring to the local dispersion
relations of the Dirac equation and the two-band semiconductor model,
respectively.
So, although the same effect happens in both systems,
this does not necessarily imply the analogy between the full underlying
Ham\-il\-to\-ni\-ans/equa\-tions of motion.

\section{Conclusions}

We studied the quantitative analogy between the Sauter--Schwinger effect and
interband tunneling in suitable semiconductors with special emphasis on
fields which depend on space and time.
To this end, we compared the Dirac Hamiltonian
{[}Eqs.~\eqref{eq:A(t)-HDirac-mom-space}
and~\eqref{eq:Phi(t,x)-HDirac-Fourier-space}{]}
in 1+1 dimensions with the effective two-band Hamiltonian of a semiconductor
{[}Eqs.~\eqref{eq:A(t)-HSC} and~\eqref{eq:Phi(t,x)-HSC-exact}{]}.
In the case of purely time-de\-pen\-dent electric fields $\Ef(t)$, one may
derive a quantitative analogy for every $k$ mode after a spatial Fourier
transform.
In this case, the analog of the Schwinger critical field~\eqref{eq:Ecrit} is
determined by material constants such as the band gap $\Eg$ and the interband
coupling $\kappa_0$, which is related to the effective mass $\meff$ via
Eq.~\eqref{eq:meff-of-kappa}.
For GaAs, for example, we obtain a value of approximately
$\Ecrit[GaAs]\approx 6.2\times 10^{8}\,\mathrm{V/m}$, which is far below the
QED critical field $\ES \approx 1.3\times 10^{18}\,\mathrm{V/m}$ and about one
order of magnitude above the typical breakdown field strength of a few
($3$--$9$) $10^7 \,\mathrm{V/m}$ in GaAs according to
Ref.~\cite{Sze+Kwok2006}.
This is a very natural result because the analog of the QED critical field
yields the ultimate quantum limit until which the semiconductor can retain its
insulating behavior:
no matter how perfect and free of defects the sample is and how low the
temperature, tunneling will become strong at that field strength (unless it is
suppressed, e.g., by a magnetic field; see below).

This scenario of purely time-de\-pen\-dent electric fields $\Ef(t)$ would
already allow us to study the analog of the dynamically assisted
Sauter--Schwinger effect~\cite{Schuetzhold+Gies+Dunne2008} with an additional
Sauter pulse, for example, where the threshold frequency
(for $\Estatic=\Ecrit[GaAs]/10$) lies around $0.12\,\mathrm{eV}$ (instead of
$80\,\mathrm{keV}$ as in real QED), which is favorable for an experimental
verification.
For the experimentally probably more relevant case of an additional sinusoidal
field (instead of a Sauter pulse), we get the additional requirement that the
field strength of this additional field should be large enough to assist
tunneling.
This indicates an important difference to the well-known Franz--Keldysh
effect~\cite{Franz1958,Keldysh1958-2} corresponding to tunneling assisted by a
single photon (which can be treated perturbatively).
A single photon with an energy of $0.12\,\mathrm{eV}$ would not have a
significant impact because its energy is far below the band gap.
However, a field oscillating at this frequency with sufficient intensity can
assist tunneling, which shows that it is necessary to treat this field beyond
(first-or\-der) perturbation theory; see also
Refs.~\cite{Linder+al2015,Torgrimsson+al2017}.

For electric fields depending on space and time, $\Ef(t,x)$, more approximations
are necessary to obtain a quantitative analogy.
For example, because electrons and positrons in real QED are limited by the
same speed of light, one has to neglect the difference in the velocities of
particles and holes (more precisely, the curvature of their bands at the gap)
in the semiconductor and to approximate both by the same effective mass of
around 7\% of the electron mass.
This scenario $E(t,x)$ includes additional interesting cases.
For example, if the strong and static field is inhomogeneous and close to the
edge of the tunneling regime, the frequency and/or field strength of the
additional weaker time-de\-pen\-dent field required for dynamical assistance
is reduced; see Sec.~\ref{subsec:dyn-ass-spacetime-dep} and
Ref.~\cite{Schneider+Schuetzhold2016-1}.

Finally, we discussed the generalization to 2+1 dimensions.
Apart from facilitating the distinction between transverse and longitudinal
fields (see also Ref.~\cite{Linder+al2015}), this case also allows us to
introduce a magnetic field.
For the Sauter--Schwinger effect in real QED, it is well known that an
additional magnetic field can suppress the tunneling probability.
Here, we find an analogous suppression for the tunneling in semiconductors; see
also Refs.~\cite{Zawadzki+Lax1966,Aronov+Pikus1967-1,Weiler+Zawadzki+Lax1967}.
For example, in GaAs with an electric field of 1\% of the critical field,
$\Estatic = \Ecrit[GaAs] / 100$ (i.e., roughly one order of magnitude below the
breakdown field strength), a magnetic field of $1\,\mathrm{Tesla}$ can already
suppress tunneling significantly
($\Estatic / \SC{\ceff}{GaAs} \approx 4.5\,\mathrm{T}$ will stop it completely).

In summary, our findings suggest that the analog of the Sauter--Schwinger effect
and its dependence on the spatial and temporal field profile (e.g., dynamical
assistance) should be observable with pres\-ent-day technology in suitable
high-qual\-i\-ty semiconductors at low temperatures, where competing mechanisms
(due to defects etc.) are suppressed sufficiently.

\section{\label{sec:outlook}Outlook: interactions}

In all of our previous considerations, we neglected the Coulomb interaction
between the electrons.
This approximation is well motivated experimentally since the picture of
non-in\-ter\-act\-ing electrons (e.g., band structure, Drude model) describes
the experiments in bulk semiconductors typically very well.
Note that the situation is different in quantum dots, for example, where the
spatial confinement enhances Coulomb interaction effects.

The same approximation is typically used in real QED, where most of the
calculations regarding the Sauter--Schwinger effect neglect the interaction
between the created electrons and positrons.
While this interaction is expected to be small, it is probably fair to say that
it is not fully understood yet.

In order to obtain a rough estimate, let us compare the Coulomb force $\FCoul$
of the elec\-tron--pos\-i\-tron pair separated by the tunneling distance to the
force $\Fext=q\Ef$ induced by the external electric field:
\begin{equation}
\frac{\FCoul}{\Fext} = \frac{1}{4} \, \alphaQED \, \frac{\Ef}{\ES} \, \text{.}
\end{equation}
Thus, even for a very strong field of $\Ef=\ES/10$, we find a suppression of
$\approx2\times10^{-4}$, which indicates that neglecting these interactions is
a good approximation.

If we now perform the same estimate for the semiconductor case, we find
\begin{equation}
\frac{\FCoul}{\Fext}
=
\frac{1}{4} \, \alphaQED \, \frac{\Ef}{\Ecrit} \, \frac{\cvac}{\ceff} \, \text{.}
\end{equation}
As a result, due to $\cvac/\ceff\approx 217$ for GaAs
(cf.\ Table~\ref{tab:A(t)-scale-subs}), the impact of the Coulomb interactions
is stronger in this situation.
Intuitively speaking, the electrons are slower and thus have more time to
interact.
This enhancement is even more pronounced for graphene~\cite{Novoselov+al2005}
where $\cvac/\ceff\approx300$.
Nevertheless, even with the very strong field $\Ef=\Ecrit[GaAs]/10$, the Coulomb
force is only a 4\% correction to the external force, such that
neglecting it should still be a good approximation.

Turning the argument around, high-pre\-ci\-sion experiments in semiconductors
could (at least qualitatively) illuminate the impact of
interactions, while the analogous experiments in real QED are far more
difficult.


\begin{acknowledgments}
The authors acknowledge financial support by the Deutsche
For\-schungs\-ge\-mein\-schaft (Grant No.\ SFB~1242, Proj\-ects~A01, B03, and
B07).
\end{acknowledgments}

\appendix

\boldmath
\section{%
\label{app:A-squared-absorption}%
Absorption of the \texorpdfstring{$A^{2}$}{A\texttwosuperior } term in the
semiconductor Hamiltonian}
\unboldmath
In the time-de\-pen\-dent case (Sec.~\ref{sec:A(t)-case}), the electric
potential is specified in temporal gauge; that is, $\Ef(t)=\dot{A}(t)$ and the
scalar potential, $\Phi$, is set to zero.
However, introducing also the scalar potential $\Phi$ explicitly for the moment,
the electric field becomes $\Ef=\dot{A}+\partial_{x}\Phi$, so a purely
time-de\-pen\-dent scalar potential $\Phi=\Phi(t)$ does not have any physical
significance.
The scalar potential couples to time derivatives
($\partial_{t}\to\partial_{t}-\I q\Phi$) and therefore leads to the additional
term $-q\Phi$ in the Hamiltonian~\eqref{eq:A(t)-HSCfull}:
\begin{equation}
\HSCfull(t)=\intop_{\mathclap{-\infty}}^{\mathclap{\infty}}
\SCFO^{\dagger}\left\{ \frac{[-\I\partial_{x}+qA(t)]^{2}}{2m}+V(x)-q\Phi\right\}
\SCFO\,\D x\text{.}
\end{equation}
We may thus absorb the quadratic $A$ term in this equation by setting
$\Phi(t)=qA^{2}(t)/(2m)$ and obtain the simplified
Hamiltonian~\eqref{eq:A(t)-HSCfull-simplified}.

\section{\label{app:mom-matrix}Bloch-wave momentum matrix elements}

\subsection{Underlying formula}

Let us first derive a general equation for a type of integrals which
appears regularly in calculations in the Bloch wave basis.
Assume that $g(x)$ is an $\LC$-pe\-ri\-od\-ic func\-tion---i.e.,
$g(x+\LC)=g(x)$---and we want to calculate the integral
$\intop_{-\infty}^{\infty}\exp(\I kx)g(x)\,\D x$ with a real $k$
satisfying $|k|<2\pi/\LC$.

We start by writing the $\LC$-pe\-ri\-od\-ic $g$ as a Fourier series,
\begin{equation}
g(x)=\sum_{j=-\infty}^{\infty}\tilde{g}_{j}\E^{2\pi\I jx/\LC}\text{,}
\end{equation}
with complex Fourier coefficients $\tilde{g}_{j}$.
Insertion into the above integral yields
\begin{align}
\intop_{-\infty}^{\infty}\E^{\I kx}g(x)\,\D x &
=\sum_{j=-\infty}^{\infty}\tilde{g}_{j}
\intop_{-\infty}^{\infty}\E^{\I(k+2\pi j/\LC)x}\,\D x\nonumber \\
 & =2\pi\sum_{j=-\infty}^{\infty}\tilde{g}_{j}
 \DiracDelta{\left(k+\frac{2\pi}{\LC}j\right)}\text{.}
\end{align}
Since $|k|<2\pi/\LC$, the delta distribution vanishes except for the case $j=0$;
cf., e.g., Ref.~\cite{Luttinger+Kohn1955}.
The corresponding Fourier coefficient, $\tilde{g}_{0}$, coincides with
the average of $g$ over a unit cell, so we get the result
\begin{equation}
\intop_{-\infty}^{\infty}\E^{\I kx}g(x)\,\D x
=\frac{2\pi}{\LC}\intop_{0}^{\LC}g(x)\,\D x\,\DiracDelta(k)\text{.}
\label{eq:periodic-func-FT-formula}
\end{equation}

\subsection{Momentum matrix elements}

We start to calculate the matrix elements by inserting the general Bloch wave
form~\eqref{eq:BW-form}:
\begin{align}
 & \braket{n,\kc|-\I\partial_{x}|n',\kc'}\nonumber \\
={} & \intop_{-\infty}^{\infty}f_{n}^{\ast}(\kc,x)
(-\I\partial_{x})f_{n'}(\kc',x)\,\D x\nonumber \\
={} & \intop_{-\infty}^{\infty}\E^{\I(\kc'-\kc)x}u_{n}^{\ast}(\kc,x)\nonumber \\
 & \hphantom{\intop_{-\infty}^{\infty}}\quad
 \times\left[\kc'u_{n'}(\kc',x)-\I
 \frac{\partial u_{n'}(\kc',x)}{\partial x}\right]\D x\text{.}
\end{align}
Since the Bloch factors are $\LC$ periodic with respect to $x$ and
$|\kc'-\kc|<2\pi/\LC$ (because $\kc$ and $\kc'$ are restricted
to the first Brillouin zone), we may apply
Eq.~\eqref{eq:periodic-func-FT-formula} and find
\begin{align}
 & \braket{n,\kc|-\I\partial_{x}|n',\kc'}\nonumber \\
={} & \Bigl[\kc\underbrace{\braketcell{n,\kc|n',\kc}}_{\delta_{nn'}}+
\braketcell{n,\kc|-\I\partial_{x}|n',\kc}\Bigr]\DiracDelta(\kc'-\kc)\text{,}
\label{eq:BW-mom-matrix-elem}
\end{align}
cf., e.g., Ref.~\cite{Gu+Kwong+Binder2013}.
Note that we used the unit-cell scalar product defined in
Eq.~\eqref{eq:Bloch-factor-orthonormalization} to write the remaining
single-cell integrals.
Furthermore, the first product just gives a Kronecker delta due to the
Bloch-factor orthonormalization~\eqref{eq:Bloch-factor-orthonormalization}.

\boldmath
\section{\label{app:kappa-Taylor-exp}%
Taylor expansion of \texorpdfstring{$\kappa(\kc)$}{kappa(K)} around $\kc=0$}
\unboldmath
We are interested in the first-order $\kc$ dependence of $\kappa$
{[}Eq.~\eqref{eq:kappa-def}{]}, so we need to evaluate the first
$\kc$ derivative of $\kappa$ at $\kc=0$.
Together with the definition of the single-cell product in
Eq.~\eqref{eq:Bloch-factor-orthonormalization}, we get ($\kc$ derivatives are
denoted as superscript numbers in parentheses)
\begin{align}
\kappa^{(1)}(0)={} &
\Braketcell{u_{\vbi}^{(1)}(0,x)|-\I\partial_{x}|u_{\cbi}(0,x)}\nonumber \\
 & {}+\Braketcell{u_{\vbi}(0,x)|-\I\partial_{x}|u_{\cbi}^{(1)}(0,x)}\text{.}
\end{align}
The $\kc$ derivatives of the Bloch factors at $\kc=0$ can be calculated
by expanding $u_{\covbi}(\kc,x)$ in powers of $\kc$ via $k\cdot p$
perturbation theory.
Again, we apply the two-band approximation, so we only take into account
corrections from the valence band and the conduction band.
The resulting expansions,
\begin{equation}
u_{\covbi}(\kc,x)=u_{\covbi}(0,x)\pm\frac{\kappa_{0}\kc}{m\Eg}u_{\vocbi}(0,x)
+\order(\kc^{2})
\label{eq:u-kp-expansion}
\end{equation}
(cf.\ Ref.~\cite{Yu+Cardona2010}), inserted above immediately give
\begin{align}
\kappa^{(1)}(0)={} &
\Braketcell{-\frac{\kappa_{0}}{m\Eg}u_{\cbi}(0,x)
|-\I\partial_{x}|u_{\cbi}(0,x)}\nonumber \\
 & {}+\Braketcell{u_{\vbi}(0,x)|-\I\partial_{x}|
 \frac{\kappa_{0}}{m\Eg}u_{\vbi}(0,x)}\nonumber \\
={} & -\frac{\kappa_{0}}{\Eg}\Delta v(0)=0
\end{align}
since the group velocities
$v_{\covbi}(\kc)=\braketcell{\covbi,\kc|-\I\partial_{x}|\covbi,\kc}/m$ vanish at
the direct band gap at $\kc=0$ in both energy bands.

The Taylor series of $\kappa$ around $\kc=0$ thus does not include
a linear term (according to $k\cdot p$ perturbation theory and the
two-band model); $\kappa(\kc)=\kappa_{0}+\order(\kc^{2})$.

\boldmath
\section{\label{app:PMSC-for-slow-Phi}%
Matrix elements of \texorpdfstring{$\PMSC(t,\kc,\kc')$}{M(t,K,K')} for spatially
slowly varying potentials}
\unboldmath
The elements of the matrix $\PMSC(t,\kc,\kc')$ in Eq.~\eqref{eq:PMSC-exact}
are expressions of the form $\braket{n,\kc|\Phi|n',\kc'}$.
For slowly varying potentials, this general scalar product can be calculated.
We start by inserting the Bloch-wave form~\eqref{eq:BW-form} and the spatial
Fourier transform {[}cf.\ Eq.~\eqref{eq:DFO-Fourier-trafo}{]} of the potential.
After changing the order of integration, we get
\begin{multline}
\braket{n,\kc|\Phi|n',\kc'}
= \frac{1}{\sqrt{2\pi}} \intop_{-\infty}^{\infty} \Phim(t,k) \\
{}\times\intop_{-\infty}^{\infty} \E^{\I(k+\kc'-\kc)x}
u_{n}^{\ast}(\kc,x)u_{n'}(\kc',x)\,\D x\,\D k\text{.}
\end{multline}

Let us now reconsider our assumptions: The slowly varying potential
satisfies $\Phim(t,k)=0$ unless $|k|\ll2\pi/\LC$, so we only need
to calculate the $x$ integral (correctly) for small values of $k$.
Furthermore, we are interested in the quasimomentum region near the
band gap to draw the analogy to Dirac theory; that is, we evaluate
the matrix elements between values of $\kc$ and $\kc'$ near the
Brillouin zone center and thus $|\kc'-\kc|$ is significantly smaller
than $2\pi/\LC$, the total zone width.
Altogether, we may assume $|k+\kc'-\kc|<2\pi/\LC$ and thus apply the formula in
Eq.~\eqref{eq:periodic-func-FT-formula} again:
\begin{multline}
\braket{n,\kc|\Phi|n',\kc'}=\frac{1}{\sqrt{2\pi}}\intop_{-\infty}^{\infty}
\Phim(t,k)\DiracDelta(k+\kc'-\kc)\\
{}\times\frac{2\pi}{\LC}\intop_{0}^{\LC}
u_{n}^{\ast}(\kc,x)u_{n'}(\kc',x)\,\D x\,\D k\text{.}
\end{multline}
Now, the $k$ integral can easily be calculated and the single-cell
$x$ integral is expressed via the Bloch-factor scalar product introduced
in Eq.~\eqref{eq:Bloch-factor-orthonormalization}.
That yields our end result
\begin{equation}
\braket{n,\kc|\Phi|n',\kc'}=\frac{\Phim(t,\kc-\kc')}{\sqrt{2\pi}}
\braketcell{n,\kc|n',\kc'}\text{.}
\end{equation}

Note that this equation is exact as long as the condition mentioned
above is true.


\input{sauter-schwinger_in_semiconductors.bbl}

\end{document}

%% file: sauter-schwinger_in_semiconductors.bbl
%